\documentclass[usenatbib]{mnras}

\usepackage{newtxtext,newtxmath}
\usepackage{bm}
\usepackage{placeins}
\usepackage{xcolor}
\usepackage[T1]{fontenc}

\DeclareRobustCommand{\VAN}[3]{#2}
\let\VANthebibliography\thebibliography
\def\thebibliography{\DeclareRobustCommand{\VAN}[3]{##3}\VANthebibliography}

\usepackage{graphicx}	% Including figure files
\usepackage{amsmath}	% Advanced maths commands
\usepackage{subcaption}
\usepackage{cuted}  % Double column equation
\newcommand{\tabascal}{\texttt{tabascal}}
\newcommand{\JAX}{\texttt{JAX}}
\newcommand{\comment}[1]{\textcolor{black}{#1}}

\title[\tabascal]{Trajectory Based RFI Subtraction and Calibration \\for Radio Interferometry}

\author[Chris Finlay et al.]{
Chris Finlay,$^{1,2,3,5}$\thanks{E-mail: christopher.finlay@unige.ch}
Bruce A. Bassett,$^{2,3,4,5}$
Martin Kunz$^{1}$
and Nadeem Oozeer$^{3,5,6}$
\\
$^{1}$D\'epartement de Physique Th\'eorique and Center for Astroparticle Physics,
Universit\'e de Gen\`eve, 24 quai Ernest  Ansermet, 1211 Gen\`eve 4, Switzerland\\
$^{2}$Department of Pure and Applied Mathematics, University of Cape Town, South Africa\\
$^{3}$African Institute for Mathematical Sciences, 6 Melrose Road, Muizenberg, 7945, South Africa\\
$^{4}$South African Astronomical Observatory, Cape Town, South Africa\\
$^{5}$South African Radio Astronomy Observatory (SARAO), 2 Fir Street, Observatory, Cape Town, 7925, South Africa \\
$^{6}$Centre for Radio Astronomy Techniques and Technologies, Department of Physics and Electronics, Rhodes University, P.O. Box 94, Makhanda 6140, South Africa
}

\date{Accepted XXX. Received YYY; in original form ZZZ}

\pubyear{2023}

\begin{document}
\label{firstpage}
\pagerange{\pageref{firstpage}--\pageref{lastpage}}
\maketitle

% Abstract of the paper
\begin{abstract}
Radio interferometry calibration and Radio Frequency Interference (RFI) removal are usually done separately. Here we show that jointly modelling the antenna gains and RFI has significant benefits when the RFI follows precise trajectories, such as for satellites. One surprising benefit is improved calibration solutions, by leveraging the RFI signal itself. We present \tabascal\ (\textbf{T}r\textbf{A}jectory \textbf{BA}sed RFI \textbf{S}ubtraction and \textbf{CAL}ibration), a new algorithm that jointly models the RFI and calibration parameters in visibilities. We test \tabascal\ on simulated MeerKAT calibration observations contaminated by satellite-based RFI. We obtain gain estimates that are both unbiased and up to an order of magnitude better constrained compared to uncontaminated data. \comment{When combined with an ad hoc RFI subtraction scheme,} \tabascal\ solutions can be further applied to an adjacent target observation: 5 minutes of calibration data results in an image with about \comment{a third} the noise achieved when using flagging alone. The recovered flux distribution of \comment{RFI subtracted} data was on par with uncontaminated data. In contrast, RFI flagging alone resulted in a higher detection threshold and consistent underestimation of source fluxes. For a mean RFI amplitude of 17 Jy, using \comment{RFI subtraction} leads to less than 1\% loss of data compared to 75\% data loss from an ideal $3\sigma$ flagging algorithm, a very significant increase in data available for science analysis. Although we have examined the case of satellite RFI, \tabascal\ should work for any RFI moving on parameterizable trajectories, relative to the phase centre, such as planes and/or objects fixed to the ground.
\end{abstract}

% Select between one and six entries from the list of approved keywords.
% Don't make up new ones.

%%%%%%%%%%%%%%%%%%%%%%%%%%%%%%%%%%%%%%%%%%%%%%%%%%

%%%%%%%%%%%%%%%%% BODY OF PAPER %%%%%%%%%%%%%%%%%%

\section{Introduction}\label{sec:intro}

A major problem plaguing radio astronomy observatories across the world is the problem of Radio Frequency Interference (RFI). In the context of radio astronomy, RFI is generally any unwanted radio signal that can result from both man-made and natural sources. The increasing sensitivity of radio telescopes coupled with more RFI sources has led to an exponentially growing number of detected sources of RFI. Several signal processing methods exist and are used to handle RFI, however, in practice there is no universal fool-proof technique for RFI mitigation. For reviews on RFI mitigation see \cite{Kesteven2010, briggs2005overview, Fridman2001, EkersRFI}.

RFI flagging, the process of identifying data points contaminated by RFI, is the most commonly used post-processing technique for RFI mitigation in use across all observatories. Though there has been significant progress in applying advances in machine learning to RFI flagging \citep{vafaei2020deep, Deep_RFI1, Deep_RFI2}, these techniques come at the expense of data loss. In this paper we instead explore statistical methods to filter out the RFI since it is reasonable to expect it to produce advantages similar to that which occurred in analysis of the Cosmic Microwave Background, see e.g. \cite{CMB_review}. In particular, Bayesian methods show significant promise for radio astronomy, e.g. \cite{BIRO, resolve3}. If successful this means losing less useful data. To do so we exploit the key defining property of RFI: namely that it is not stationary in the reference frame of the celestial sky. RFI is always moving relative to the sky, either due to the earth's rotation or artificial orbits (planes and satellites).

Methods for the subtraction of RFI signal have been investigated in the past with limited success. \cite{perley2003removing} proposed a method for subtracting RFI visibility contributions. In \cite{cornwell2004rfi} the proposed method was tested on carefully chosen real data and managed to reduce the RFI contribution by up to a factor of 1000 in specific channels of a ground-based RFI source. This method therefore showed real promise and appears to have resulted in a task named \texttt{UVRFI} \citep{kogan2010evla} that is available in the \comment{Astronomical Image Processing System (AIPS) software \citep{greisen2002aips}}. The \texttt{UVRFI} task has two sub tasks: (1) \texttt{CIRC}, an extension of RfiX \citep{athreya2009new}, that is closely related to the original method, is used for ground-based sources and fits a linearly varying amplitude given the implied fringe-frequency, and (2) \texttt{CEXP}, that applies \texttt{CLEAN} \citep{clean} in the Fourier domain of the visibility time series, \comment{that can be used for any RFI source as long as its fringe frequency is suitably different from that of the celestial signal and the RFI visibility amplitude has not been fringe-washed away.}

Our proposed method, \tabascal, has some minor similarities with the \texttt{CIRC} method from \texttt{UVRFI}. However, it is more general as it includes the ability to work on both stationary and moving sources of RFI, given that the RFI source moves on a fixed trajectory we can parameterize. This is the case for all stationary sources as well as most moving sources such as planes and satellites. The key differences are (1) the full decomposition of $N(N-1)/2$ baseline signals into $N$ antenna-based signals, (2) the modelling of time-smearing (fringe winding loss) effects on the signal, and (3) the joint estimation of antenna gains and RFI parameters. In this paper we apply \tabascal\ to simulations of MeerKAT \citep{jonas2016meerkat} observations contaminated by satellite-based RFI.

We have chosen this situation as a testbed as MeerKAT is one of the most sensitive radio telescopes in the world, in its operational bands, and is the precursor for the Square Kilometre Array (SKA) Mid telescope. Additionally, its L-band (900 - 1670 MHz) is already severely affected by satellite-based RFI, not to mention the increasing number of satellites yet to be deployed, such as the SpaceX Starlink (10.7 - 12.7 GHz) and Amazon Kuiper constellations, which will affect radio telescopes in a number of frequency bands. The MeerKAT L-band data currently suffers around \comment{35\%} data loss due to RFI flagging \citep{sihlangu2021multi} of which the majority is caused by satellite-based sources. Furthermore, satellites are a particularly interesting source of RFI from the perspective of RFI subtraction due to their predictability. The positions of satellites can be predicted based on previous measurements, such as from two-line element sets (TLEs), and for many, the spectral and temporal signal profiles are known a priori \citep{harper2018potential}. \comment{However, in our analysis, the only a priori information about the RFI signal assumed is that it obeys amplitude closure at our chosen sampling frequency.} The identified satellite constellations that currently affect MeerKAT L-band data are summarised in Table \ref{tab:sats}.

\begin{figure}
  \centering
  \includegraphics[width=0.95\linewidth]{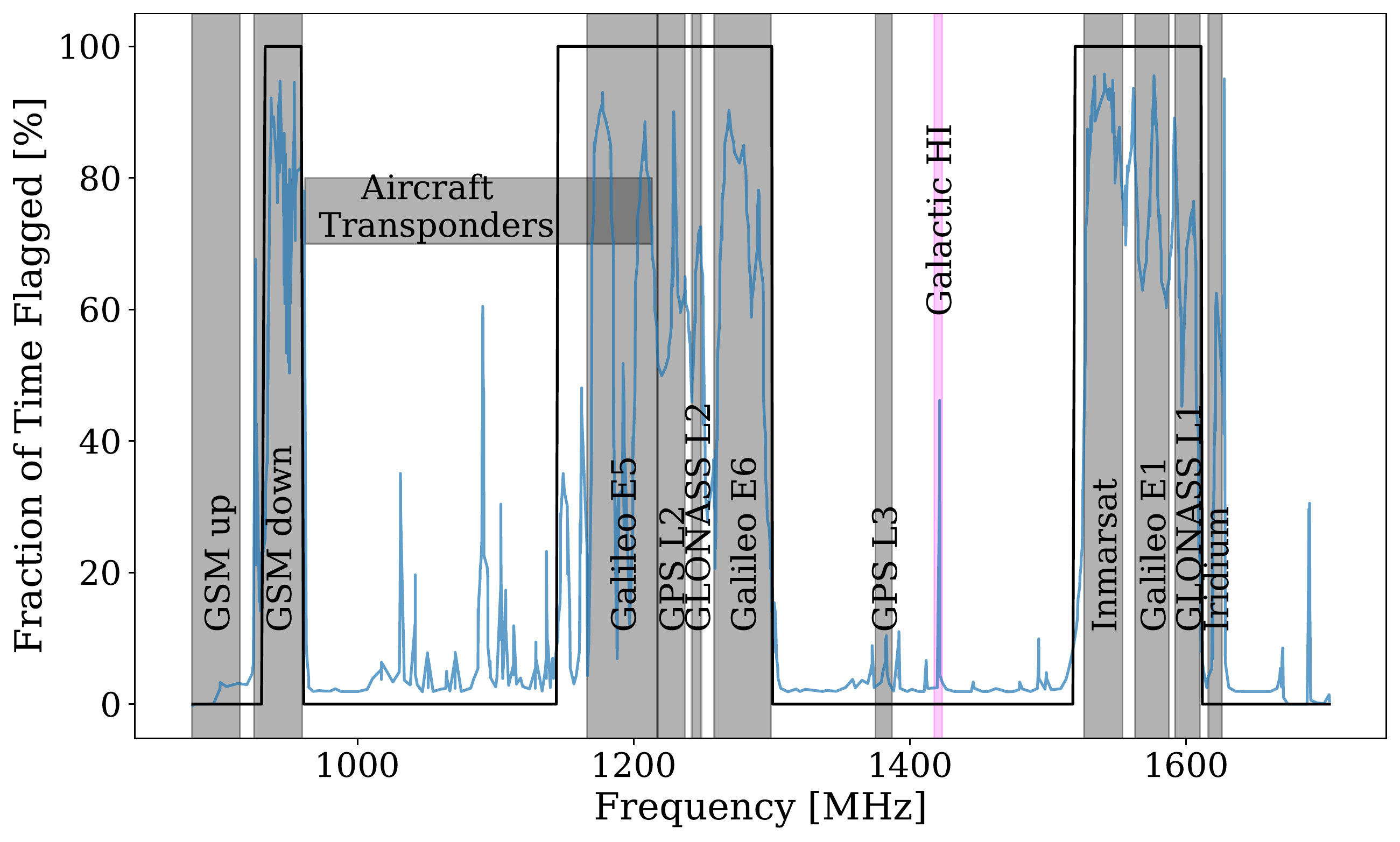}
  \caption{\comment{The fraction of time that is flagged by the MeerKAT RFI flagger in the L-band \citep{sihlangu2021multi}}. The most heavily affected sub-bands have a static mask (black line) on the baselines shorter than 1 km. We can see that the majority of the RFI present in this band is from satellite-based RFI of which the Global Navigation Satellite System (GNSS) are the main culprit.}
  \label{fig:rfi_lband}
\end{figure}

\begin{table}
  \begin{center}
    \begin{tabular}{|c|c|c|c|}
      \hline
      Constellation & Frequency     & Orbit       & Orbit  \\
                    & Bands (MHz)   & Elevation   & Inclination \\
      \hline
      GPS     &   \begin{tabular}{@{}c@{}} L1: 1565 - 1585 \\
                                           L2: 1217 - 1237 \\
                                           L3: 1375 - 1387 \\
                                           L5: 1166 - 1186 \\
                  \end{tabular}                                   & 20,180 km  & 55$^\circ$ \\
      \hline
      GLONASS &   \begin{tabular}{@{}c@{}} L1: 1592 - 1610 \\
                                           L2: 1242 - 1249 \\
                                           L3: 1202.025    \\
                  \end{tabular}                                   & 19,140 km  &  64.8$^\circ$ \\
      \hline
      Galileo &   \begin{tabular}{@{}c@{}} E1:   1575.42       \\
                                           E5a: 1176.45        \\
                                           E5b: 1207.14        \\
                                           E5 AltBOC: 1191.795 \\
                                           E6: 	1278.75        \\
                  \end{tabular}                                   & 23,222 km  &  56$^\circ$ \\
      \hline
      Inmarsat &  1526 - 1554                                     & 35,785 km  &  0$^\circ$  \\
      \hline
      Iridium  &  1616 - 1626                                     & 780 km     & 86.4$^\circ$ \\
    \end{tabular}
  \end{center}
  \caption{A summary of the characteristics of the satellite-based RFI that affects the MeerKAT L-band. All of the active satellites in these constellations have eccentricities of less than 1 \%.
  Sources: \href{https://gssc.esa.int/navipedia/index.php/GLONASS_Space_Segment}{GLONASS ESA},
           \href{https://novatel.com/an-introduction-to-gnss/chapter-3-satellite-systems/glonass-global-navigation-satellite-system-russia}{GLONASS Novatel},
           \href{https://gssc.esa.int/navipedia/index.php/GPS_Space_Segment}{GPS ESA},
           \href{https://gssc.esa.int/navipedia/index.php/Galileo_Space_Segment}{Galileo ESA},
           \href{https://skaafrica.atlassian.net/wiki/spaces/ESDKB/pages/305332225/Radio+Frequency+Interference+RFI}{SARAO RFI Report}.
           \href{https://earth.esa.int/web/eoportal/satellite-missions/i/iridium-next}{Iridium ESA}}
\label{tab:sats}
\end{table}
\FloatBarrier

This paper is organized as follows: Section \ref{sec:method} describes the data simulations, the probabilistic inverse model and the methods used to recover parameters of interest. In Section \ref{sec:results} we discuss the recovered posterior parameter distributions and how these results are used to improve science performance in a target observation. Finally, in Section \ref{sec:conclusions} we summarize the problem and our findings as well as further directions for this work.

\section{Methods and Simulations}\label{sec:method}

This section is organised as follows: in Sections \ref{sec:interferometry} \& \ref{sec:response} we discuss the principle of radio interferometry and how we model the response of the telescope to the sky brightness distribution. In Sections \ref{sec:rfi_mod} \& \ref{sec:data_gen} we discuss modifications to the standard model necessary for most types of RFI and the specific implementation of our MeerKAT data simulations. Sections \ref{sec:bayes}, \ref{sec:likelihood} \& \ref{sec:prob_model} start with a brief introduction to Bayesian concepts used in this paper, and then go on to describe our likelihood term and forward model along with the associated priors. We finish off with Sections \ref{sec:optimize} \& \ref{sec:MCMC} where we describe the methods used to recover the posterior distributions of our model parameters by means of a Laplace approximation and Markov chain Monte Carlo (MCMC).

We begin by describing the problem of transfer calibration, also referred to as $1^\text{st}$ Generation Calibration (1GC). 1GC is the first step in the calibration of an interferometer that provides a starting point for further calibration i.e. 2GC/\texttt{selfcal}. In 1GC, observations of a calibrator source, i.e. a source of known position and flux, is used to estimate the antenna gains which can then be used do an initial calibration on a following target observation where the sky distribution is not known a priori. \comment{A suitable calibrator source is unresolved on the longest baseline, or at least only partially resolved, and whose flux density is comparable to the array system equivalent flux density (SEFD). It should also be significantly brighter than all other sources in the field of view (FoV), if this is not true, then a source model is required for the other sources as well.} To also serve as a suitable phase calibrator it should be close to the target field (within 10$^\circ$ on the sky) \citep{mauch20201}, this is defined by the angular scale of \comment{atmospheric and ionospheric fluctuations} which affect the gain phases. \comment{The brighter the calibrator source, the greater the signal-to-noise ratio (SNR) achieved (until the limits of the instrumental/atmospheric perturbations are reached), leading to better constrained antenna gain estimates for a given observation time.} A list of suitable calibrators is available from the SARAO External Service Desk\footnote{\href{https://skaafrica.atlassian.net/wiki/spaces/ESDKB/overview}{https://skaafrica.atlassian.net/wiki/spaces/ESDKB/overview}}.

\subsection{Radio Interferometry}\label{sec:interferometry}

A radio interferometer measures the sky brightness distribution (spectral radiance) by \comment{measuring the spatial coherence of the signal. This is known as visibility space and is closely related to the Fourier space of the sky brightness.} The domain of visibility space is denoted by the coordinates $(u,v,w)$. To sample a point in visibility space, the signal from two antennas with some spatial separation, measured in wavelengths, is correlated. The separation vector, given by the components $(u,v,w)_{pq}$, referred to as a baseline, points from antenna $p$ to antenna $q$. This is done for all antenna pair combinations in an interferometric array at multiple time steps leading to samples of many locations in the visibility space. One is then able to infer the sky brightness distribution from the visibility samples using the formalism described in this section.

For an ideal \comment{fringe-stopping interferometer}\footnote{\comment{A fringe-stopping interferometer introduces time/phase delays in the signal chain such that the fringe pattern remains stable in the direction of the phase centre.}}, the visibility distribution is defined by,
 \begin{equation}\label{eq:ideal_vis}
  V(u,v,w) = \iint\limits_{lm} I(l,m) \exp \bigg[ -2\pi i \big(ul + vm + w(n-1) \big) \bigg] \frac{dldm}{n},
\end{equation}
where $i$ is the imaginary unit and $I(l,m)$ is the brightness distribution on the sky. The sky coordinates, $(l,m,n)$, are unitless direction cosines lying in the range $[-1,1]$. Since the domain of the sky brightness distribution, $I$, lies on a sphere of fixed arbitrary radius, the celestial sphere, the third direction cosine, $n$, is fixed, i.e. $n=\sqrt{1-l^2-m^2}$. The origin of the sky coordinates, $(0,0,1)$, is called the phase centre and is fixed to a location on the celestial sphere. This is the defining property of a fringe-stopping interferometer. When considering the field of view (FoV) of the telescope to be very small, i.e. the sky signal is dominated by an area of the sky where $l^2+m^2 \ll 1$, Equation \eqref{eq:ideal_vis} reduces to the van Cittert-Zernike (vCZ) theorem.
In this case, $n\approx1$, producing an exact 2D Fourier relation between the visibilities and the sky brightness distribution \citep[][Chapter 15.1.1]{thompson2017interferometry}.

\subsection{Telescope Response}\label{sec:response}

The telescope response, $\Gamma_{pq}$, to measuring some sky brightness distribution includes the modulation of the primary beam of each of the antennas $p$ and $q$ and their respective bandpass filters. When the primary beam intensity pattern of antenna $p$ is $|E_p(l,m,\nu,t)|^2$ and its bandpass is $|G_p(\nu,t)|$, the telescope response is as given in Equation \eqref{eq:tel_response1}. This is very similar to the form given in \citealt[][Chapter 3.3.4]{thompson2017interferometry}.

\begin{equation}\label{eq:tel_response1}
    \Gamma_{pq} = \frac{1}{\Delta t}\iint\limits_{\begin{subarray}{l}
                              t_j \pm \Delta t/2 \\ \nu_0 \pm \Delta\nu/2 \end{subarray}}
        G_p G_q^* \iint\limits_{lm} E_p E_q^* I \exp \left[ ... \right] \frac{dldm}{n} dt d\nu
\end{equation}

In Equation \eqref{eq:tel_response1}, $^*$ indicates the complex conjugate, $\Delta t \gg \Delta \nu^{-1} $ is the integration time in the correlator for a single sample at time $t_j$ and $\Delta \nu$ is the bandwidth of a single frequency channel centred on $\nu_0$. We have emitted the arguments of $G$, $E$ \& $I$ but in principle all of these depend on $\nu$ and $t$ and are generally assumed constant over the intervals $\Delta\nu$ and $\Delta t$.
Only $E$ and $I$ are functions of $l$ and $m$. The expression inside the exponential denoted by $[...]$ is the same as in Equation \eqref{eq:ideal_vis}.
Once the integrals are calculated we are left with Equation \eqref{eq:tel_response2}

\begin{equation}\label{eq:tel_response2}
  \Gamma_{pq}(u,v,w) = |G_p||G_q|e^{i(\varphi_p-\varphi_q)} \Delta\nu \sqrt{A_p A_q} V_{pq}(u,v,w)
\end{equation}

where $V_{pq}$ is the true visibility at the point $(u,v,w)_{pq}$, $A_p$ is the collecting area of the dish on antenna $p$ in the direction of the phase centre, $|G_p|$ is the magnitude of the bandpass/gain at antenna $p$ and $\varphi_p-\varphi_q$ is the phase difference of the gains between antennas $p$ and $q$.

In Equation \eqref{eq:tel_response2} we have assumed that the gains are constant over the integration time and the channel bandwidth. This is a standard assumption as telescopes are designed to have such stability. Additionally, due to the rotation of the Earth and frequency dependence of the $(u,v,w)$ coordinates, the visibility phases (from the exponential term in Equation \eqref{eq:ideal_vis}) vary over time and frequency. This variation cause the phases to change slightly over the integration window leading to decorrelation of the signal. This results in a reduction in the amplitude of the visibilities. This effect is known as time/frequency smearing. \comment{The strength of this effect increases with baseline length and the offset of a source from the phase centre.} Therefore, integration windows and frequency channels are chosen to be small enough to minimize this effect for astronomical sources. Typically, the integration windows are still too large for signals from RFI sources to correlate fully. This is treated as a feature to reduce the level of contamination. The telescope response $\Gamma_{pq}$ is in units of Watts when the sky brightness distribution, $I(l,m)$, is in units of W.m$^{-2}$.Hz$^{-1}$.

% Maybe the units of I should be should be W.m$^{-2}$.Hz$^{-1}$.sr$^{-1}$ but I have removed .sr$^{-1}$ due to the conversion of the integral from $d\Omega$ to $dldm/n$.

When modelling the visibilities we discretize the sky brightness distribution as a collection of point sources which are summed over, as in Equation \eqref{eq:RIME}.

\begin{equation}\label{eq:RIME}
  \tilde{V}_{pq} = G_p \left( \sum_{s} E_{ps} K_{ps} B_s K^*_{qs} E^*_{qs} \right) G^*_q
\end{equation}

where the measured visibility, $\tilde{V}_{pq}=\Gamma_{pq}$, the $E$ terms are normalized as $E_p \rightarrow E_p/\sqrt{A_p}$ and the $G_p \rightarrow G_p/\sqrt{\Delta \nu}$ making them dimensionless in the equation. This way we have both the point source brightnesses, $B_s$, and the measured visibilities, $\tilde{V}_{pq}$, in the same units, Jansky (Jy). Those familiar with the Radio Interferometry Measurement Equation (RIME) \citep{RIME2} will recognize Equation \eqref{eq:RIME}, however, here we use it in a scalar form and do not consider polarization.
In Equation \eqref{eq:RIME} we use the subscript $s$ to label a point source at position $(l_s,m_s,n_s)$ in the sky.

In Equation \eqref{eq:RIME} the $K$ terms, as defined in Equation \eqref{eq:geo_delay_far},

\begin{equation}\label{eq:geo_delay_far}
  K_{ps} = \exp \left(- 2\pi i \left(u_pl_s + v_pm_s + w_p(n_s-1) \right) \right)
\end{equation}

are the geometric delay between an antenna and an arbitrary reference position. Their combination $K_{ps} K_{qs}^*$ produce the exponential term in Equation \eqref{eq:ideal_vis} for a specific location $s$. The $B_s$ term is the spectral flux density of the point source $s$ in units of Jy. Extended sources are represented by a discretized/pixelized version where each pixel is treated as a point source.
The $E$ terms are the direction-dependent effects (DDEs), previously used for the primary beam in Equation \eqref{eq:tel_response1}. The $G$ terms are the direction-independent effects (DIEs), previously used for the gains in Equation \eqref{eq:tel_response1}.

In Equation \eqref{eq:geo_delay_far} the $(l,m,n)_s$ are the coordinates of the point source $s$ and $(u,v,w)_p$ are the coordinates of antenna $p$ relative to a global reference position, in units of wavelength.

\subsection{Modifications for RFI}\label{sec:rfi_mod}

A number of features make signals from RFI sources distinct from astronomical sources. RFI sources are much closer to us than astronomical sources and move relative to the celestial sphere on which astronomical sources, outside the solar system, remain stationary. We will start by discussing the implications of receiving signal from a source closer than expected.

Spherical waves, originating from a source, resemble plane waves when the source is very far away from the receiver, relative to the receiver dimensions and wavelength of the emission. This is known as the far-field regime and is defined in Equation \eqref{eq:far-field},
\begin{equation}\label{eq:far-field}
  d_F \gg \frac{2D^2}{\lambda}, \quad \text{given} \quad d_F \gg D \quad \& \quad d_F \gg \lambda,
\end{equation}
\comment{where $\lambda$ is the observation wavelength, $d_F$ is the distance between the emitter and receiver and $D$ is the diameter of the receiving dish (for a single dish) or baseline (for an interferometer).} For a single MeerKAT receptor observing in the L-band ($\lambda \approx 21$ cm), $d_F$ is just over 2 km which is also much larger than the dish diameter, $D$, of $\approx 14$ m. Therefore, the far-field primary beam models used for astronomical sources are also suitable for use with most sources of non-local RFI, especially, satellite-based sources.

The geometric delay term, given by Equation \eqref{eq:geo_delay_far}, that is used in the RIME, assumes sources are in the far-field of the antenna array, not just a single receiver. For the MeerKAT telescope, observing in the L-band,  with the longest baseline of $\approx 8$ km, $d_F$ is $\approx 6.4\times10^5$ km which is nearly 2 times the distance from the Earth to the Moon. Therefore, practically all sources of man-made RFI are in the near-field of the telescope array. For such sources, Equation \eqref{eq:geo_delay_far} must therefore be modified to Equation \eqref{eq:geo_delay_near}:
\begin{align}\label{eq:geo_delay_near}
    K_{ps}(t) = \exp \Bigg( -2\pi i \bigg( \frac{|\vec{r}_s(t) - \vec{r}_p(t)|}{\lambda} - w_p(t) \bigg) \Bigg)
\end{align}
where we have assumed spherical wave fronts as can be expected from a point source in the idealised case. In Equation \eqref{eq:geo_delay_near}, $\lambda$ is the observation wavelength, $\vec{r}_s$ is the position of the RFI source, $\vec{r}_p$ is the antenna position, and $w_p$ is the phase tracking delay correction as used in Equation \eqref{eq:geo_delay_far}.
When combined as a $K_p K_q^*$ term, Equations \eqref{eq:geo_delay_far} \& \eqref{eq:geo_delay_near} describe the same thing, the path length difference between antennas $p$ and $q$, in wavelengths, including the phase tracking correction.

For a far-field source (relative to the array size), the angle, relative to the reference direction, at which the signal enters the primary beam is the same across all antennas. However, this is not the case for near-field sources. \comment{In this case, the angular separation between the pointing direction and the RFI source is different for each antenna, leading to separate primary beam modulations ($E$ terms) per antenna, for a given RFI source. In our simulations, the functional form of the $E$ terms are identical across antennas. In reality this would not be the case, especially in the sidelobes. Fortunately, our subtraction method is unaffected by this fact as we model the product of the RFI signal and primary beam modulations per antenna and time step without any assumption of a primary beam model.}

Another consideration for near-field sources is the the spectral flux density received at the antenna. The spectral flux density, $I_{ps}^\text{RFI}$, of an RFI source $s$ at antenna $p$ is given by
\begin{equation}\label{eq:rfi_intensity}
  I_{ps}^\text{RFI}(t) = \frac{P_\nu^\text{RFI}(t)}{4 \pi |\vec{r}_s(t) - \vec{r}_p|^2} .
\end{equation}

Equation \eqref{eq:rfi_intensity} \citep{perley2002attenuation} is derived from the free-space path loss formula assuming spherical wave fronts from an isotropic emitter. In reality, an RFI source will not be an isotropic emitter however this would only change the $P_\nu^\text{RFI}(t)$ term making it antenna dependent. \comment{This is not a problem for our method due to the per antenna parameterization used in our forward model for the subtraction, Section \ref{sec:RFI_recovery_model} describes this in further detail.}

In Equation \eqref{eq:rfi_intensity}, $P_\nu^\text{RFI}$ is the spectral flux of the RFI source in W.Hz$^{-1}$ which can be divided by $10^{26}$ to make the spectral flux density in units of Jy, assuming the distance is in metres. Since this is different for each antenna we adapt the $B_s$ term in Equation \eqref{eq:RIME} to a $B_{pqs}$ term as defined in Equation \eqref{eq:rfi_B}.

\begin{equation}\label{eq:rfi_B}
  B_{pqs}^\text{RFI} = \sqrt{I_{ps}^\text{RFI}I_{qs}^\text{RFI}}
\end{equation}

Finally, to address the moving nature of RFI sources relative to celestial sources, we need to explicitly perform the integration for each visibility sample. Typically for radio interferometry simulations, one can evaluate all terms in the model only once per visibility sample as everything is assumed constant on the time-scale of the visibility sample integration time. \comment{However, this assumption breaks down for nearly all RFI sources due to their fringe frequency and movement through the antenna sidelobes. Therefore, we must evaluate the RFI visibilities at a finer time resolution and then average them to the cadence of the final visibility samples. Under the assumption the amplitude of the RFI signal is constant over the integration window at the finer time resolution and no nonlinearities are present, this will then account for the fringe winding loss and appropriately averaged visibility phase. The fringe frequency of a source $s$ is given by the time derivative of the instantaneous visibility phase divided by $2\pi$, where the visibility phase is the argument of $K_{ps} K_{qs}^*$ on baseline $pq$.}

\comment{Under the above conditions, the visibility amplitude accounting for fringe winding loss, $|V^\text{avg}|$, over an integration time $\Delta t$ is given by:
\begin{equation}\label{eq:RFI_winding_loss}
    |V^\text{avg}| = |V^\text{inst}| \text{ sinc} \left( \nu_f \Delta t \right),
\end{equation}
where $|V^\text{inst}|$ and $\nu_f$ are the instantaneous visibility amplitude and fringe frequency of the source respectively \citep{perley2002attenuation}. The error introduced by assuming the amplitude of the visibility is constant in a finite time averaging window leads to a non-closing error. \cite{perley2003removing} derives a lower bound for the sampling frequency, $\nu_s=1/\Delta t$, required to reduce the closure error below the noise level and is given by:
\begin{equation}\label{eq:RFI_sampling_rate}
    \nu_s > \pi \nu_f \sqrt{\frac{|V_\text{inst}|}{6 \sigma_n}},
\end{equation}
where $\sigma_n$ is the noise in the visibility data sample. Given the trajectory of the RFI source is reasonably well known a priori, the fringe frequency on any baseline for a given pointing can be estimated. The instantaneous visibility amplitude can be estimated from specific baselines where the fringe rate is slow enough such that there is negligible fringe winding loss. These specific baselines will typically be the shortest however the orientation of the baseline relative to the direction of motion of the RFI source matters.} 

\comment{The $\Delta t$ used in Equations \eqref{eq:RFI_winding_loss} \& \eqref{eq:RFI_sampling_rate} are the finer time resolution referred to in the two paragraphs above and not the final integration time of the visibility data. However, Equation \eqref{eq:RFI_winding_loss} applies at all integration lengths. Therefore, for a long enough integration time and/or high enough fringe frequency the RFI visibility will have been washed out in the averaging process. At this point the RFI visibility signal could no longer be used as a calibrator as is described in this paper, however, fortunately it should no longer pose a problem to the traditional calibration process.}

\subsection{Data Generating Model}\label{sec:data_gen}

In this section we describe the data generation model. We split this section into the telescope, DIE, DDE, astronomical and RFI source models. Together these components fully define our model used to simulate a calibration observation from the MeerKAT telescope with basic, but, realistic telescope response, signal corruptions and RFI contamination. While we choose to simulate MeerKAT observations for illustrative purposes, everything in our algorithm applies to other radio interferometry observatories. 

Table \ref{tab:data_gen_params} summarizes the parameter values and distributions used. Throughout our simulations we work with only a single frequency channel centred at 1.227 GHz and calculate observed visibilities using Equation \eqref{eq:RIME}. Each time sample, $t_j$, is an average over a given integration time, $\Delta t=$ 2s, where the visibility function is sampled 16 times per second. These integration samples are equally spaced in the range $(t_j-\Delta/2 t, t_j+\Delta t/2)$ where $t_j$ is the observation time centroid of the time sample and $\Delta t$ is the integration time.

\subsubsection{Telescope Model}

We simulate data for the MeerKAT telescope positioned at a latitude, longitude and elevation of $(-30.721^\circ, 21.411^\circ, 1054.71\text{m})$. We use the East, North, Up (ENU) coordinates obtained from the South African Radio Astronomy Observatory (SARAO) to simulate $(u,v,w)$ coordinates. We use standard coordinate transformations, as defined in \citealt[][Chapter 4.1]{thompson2017interferometry}, to transform from ENU coordinates to International Terrestrial Reference Frame (ITRF) coordinates and then to $(u,v,w)$ coordinates.

\subsubsection{Astronomical Source Model}

We simulate both a calibration observation and a target observation. In the calibration portion we observe a 1 Jy point source situated at $(\alpha,\delta)=(21^\circ,10^\circ)$. \comment{We include no other astronomical sources in the calibration portion and assume the source flux density and position is known perfectly in our RFI subtraction analysis.} In the target portion, we observe a 100 point source field, centred on $(\alpha,\delta)=(27^\circ,15^\circ)$, where the positions are uniformly sampled from a disk with 0.5$^\circ$ radius and the spectral flux densities sampled from an exponential distribution with mean 0.1 Jy. The source positions are chosen such that no two sources are closer than $\approx 8$ synthesized beam widths (80"). The simulated visibilities for the calibration portion will be denoted by $V^\text{CAL}$, and for the target track we will use $V^\text{AST}$.

\subsubsection{RFI Source Model}\label{sec:rfi_model}

Our satellite-based RFI source is modelled with a spectral flux, $P_\nu^\text{RFI}$, that is constant in time. This is done for simplicity and is not a requirement for the functioning of our method as we allow for a time variable RFI amplitude. We use the near-field expression for the spectral flux density at a specific antenna $p$ as defined in Equation \eqref{eq:rfi_intensity}. This leads to the baseline dependent spectral flux density $B_{pqs}$ as defined in Equation \eqref{eq:rfi_B}.

The position of the satellite-based RFI source is modelled using a circular orbit about the Earth. One could use a more sophisticated model at the expense of introducing more parameters. Simplified perturbation models, such as SGP4/SDP4 \citep{sgp4} would be the preferred model to use, however, since we are currently testing on simulated data with satellites with very low eccentricity we decided on a simpler model. For use on real data a more sophisticated model may very well be needed. For a circular orbit we have four parameters, namely, the orbit elevation ($h$), argument of perigee ($\gamma$), orbit inclination ($\beta$), and the right ascension of the ascending node ($\tilde{\alpha}$). The formula for circular motion on an arbitrary plane about the Earth's centre of mass, $\vec{r}_\text{e CoM}$ as a function of time is given in Equation \eqref{eq:rfi_pos} \citep{fitzpatrick2012introduction}.

\begin{equation}\label{eq:rfi_pos}
  \vec{r}_{RFI}(t) = \bm{R}_z(\tilde{\alpha})\bm{R}_x(\beta)\bm{R}_z(\gamma)
  \begin{pmatrix}
    (R_e+h)\cos(\omega t) \\
    (R_e+h)\sin(\omega t) \\
        0
  \end{pmatrix}
  + \vec{r}_\text{e CoM}(t)
\end{equation}

In Equation \eqref{eq:rfi_pos} above $\bm{R}_x(\beta)$ is a 3D rotation matrix about the $x$-axis through an angle $\beta$, $h$ is the orbit elevation above the Earth's surface in metres, $\beta$ and $\tilde{\alpha}$ define the orbital plane, $\gamma$ is the angular offset of the orbit and finally $\omega=\sqrt{G_0M_e/(R_e+h)^3}$ is the angular speed of the satellite. Here we have assumed the satellite's mass to be negligible compared to the mass of the Earth. In the equation for $\omega$, $M_e$ and $R_e$ are the mass and average radius of the Earth respectively and $G_0$ is the gravitational constant. The angular orbit parameters align with orbital elements used in two-line element sets (TLE). The orbit elevation, for a circular orbit, is directly related to the mean motion orbital element, in revolutions per day, by $86400\omega/2\pi$. The rotation matrix axes are with respect to an Earth-centred inertial (ECI) frame where $+z$ points from the Earth's centre of mass to the North Pole and $+x$ points to the vernal equinox.

The time averaging described in the beginning of Section \ref{sec:data_gen} is a fundamental requirement in modelling RFI visibility contributions. This is because the visibility phases induced by an RFI source are rapidly varying in time due to their fast movement relative to the sky reference frame. The resulting effect is called time-smearing (fringe winding) and is especially prominent for moving sources. The magnitude of this effect increases with the length of the baseline and therefore affects longer baselines more than shorter baselines, assuming the same orientation. \comment{For our simulation of a GPS satellite and the MeerKAT telescope the fastest fringe-rate is 1.3 Hz. Given the maximum instantaneous RFI visibility amplitude is 43 Jy and the visibility noise is 0.65 Jy in our simulation, using Equation \eqref{eq:RFI_sampling_rate} we obtain a minimum sampling rate of 13.7 Hz. We have therefore used 16 Hz. Since certain baselines have fringe-rates very close to 0.5 Hz and 1 Hz and our integration time for a data sample is 2 s, the RFI phase wraps an integer number of times. On these baselines the RFI signal has been averaged away nearly completely leading to orders of magnitude in RFI suppression as described by Equation \eqref{eq:RFI_winding_loss}. We also have baselines on which the fringe-rate is nearly 0 Hz and therefore no fringe winding loss has occurred.}

\subsubsection{Direction Independent Effects (DIE)}

We include time-varying complex gains for each antenna. Both the gain amplitudes and phases are modelled as linear time variates, as shown in Equations \eqref{eq:gain_model}.

\begin{subequations} \label{eq:gain_model}
  \begin{align}
    |G|(t) &= |G|^{(0)} + \dot{|G|}t \label{eq:gain_amp_model} \\
    \varphi_{G}(t) &= \varphi_{G}^{(0)} + \dot{\varphi}_{G}t \label{eq:gain_phase_model} \\
    G(t) &= |G|(t)\exp \big[ i \varphi_{G}(t) \big]
  \end{align}
\end{subequations}

The initial values and rates of change are sampled from the distributions described in Table \ref{tab:data_gen_params} and is done separately for each antenna.

$|G|(t)$ is the gain amplitude and $\varphi_{G}(t)$ is the gain phase. In Equations \eqref{eq:gain_model}, when the parameter has a $(0)$ superscript, it is the initial value, at $t=0$, and the overdot is used for the rate of change of the parameter. The gain phases include the ionospheric effects, a DDE component, but the spacial scale of variation is assumed to be so large $(>10^\circ)$ that it can be considered a DIE. We therefore only consider our RFI source to be within this angular distance from the pointing centre for this approximation to be valid. To extend our method to such a situation we could explicitly include the ionospheric effects in the DDE term. This would further require our forward model to be correspondingly adapted. 

\subsubsection{Direction Dependent Effects (DDE)}

We use the normalised Fourier transform of a circular aperture, the square of which is the normalized Airy disk, as the primary beam voltage model. The primary beam is the only DDE that we include. We keep the primary beam constant in time and the same across all antennas. \comment{This is only done in the data simulation but is not assumed in the subtraction analysis.} The functional form of the primary beam term is given by

\begin{equation}\label{eq:beam_model}
    E(\theta, \nu) = \frac{2 c_0 J_1 \big( \pi D \nu \sin\theta /c_0 \big)}{\pi D \nu \sin\theta},
\end{equation}
where $J_1$ is the Bessel function of the first kind of order one, $\nu$ is the observation frequency in Hz, $D$ is the dish diameter, $\sin\theta=\sqrt{l^2+m^2}$ where $\theta$ is the angular separation between our pointing direction and the source and $c_0$ is the speed of light in a vacuum.

We only consider a real-valued primary beam voltage model and leave complex voltage patterns and other DDEs for further study. The inclusion of complex DDEs, such as a complex voltage pattern and ionospheric effects, create a degeneracy in the forward model that would need to be broken by the inclusion of appropriate priors in the probabilistic model.

\subsubsection{Noise Model}\label{sec:noise}

We add circularly-symmetric complex normally distributed noise to the visibilities as defined in \citealt[][Chapter 6.2.2]{thompson2017interferometry}. Each baseline is  an independent measurement with independent noise. The standard deviation, $\sigma_n$, of the noise is the same for each baseline. Therefore, our modelled visibility data are independent and identically distributed (i.i.d.). The noisy data is generated using Equation \eqref{eq:RIME_noise}, where $\eta_{pq}$ is the noise term.

\begin{equation}\label{eq:RIME_noise}
  \hat{V}_{pq} = G_p \Big( \sum_{s} E_{ps} K_{ps} B_s K^*_{qs} E^*_{qs} \Big) G^*_q + \eta_{pq}
\end{equation}

In \cite{jonas2016meerkat}, measurements show that the System Equivalent Flux Density (SEFD) of a single MeerKAT receptor is approximately 420 Jy at 1.227 GHz. Using 
\begin{equation}\label{eq:SEFD}
  \sigma_n = \frac{\text{SEFD}}{\sqrt{\Delta \nu \Delta t}},
\end{equation}
this implies a per visibility noise level, $\sigma_n$, of about 0.65 Jy using a 2 s integration time, $\Delta t$, and 209 kHz bandwidth, $\Delta \nu$. We therefore model the noise as $\mathcal{CN}(0,0.65^2)$ which is equivalent to $\mathcal{N}\left(0, 0.65^2/2\right)$ in both the real and imaginary parts of the visibility independently.

\subsubsection{Summary of Parameter Values}

\def\arraystretch{1.2}%  1 is the default, change whatever you need
\begin{table}
\begin{center}
  \resizebox{0.45\textwidth}{!}{%
  \begin{tabular}{ |cc|cc|cc|cc| }
    \hline
    Parameter Description & Symbol & Units & Value/Distribution \\
    \hline
    \hline
    Dish Diameter                   & $D$                    & m                     & $13.965$                        \\
    Observation Frequency           & $\nu_0$                & GHz                   & $1.227$                           \\
    Channel Bandwidth               & $\Delta \nu$           & kHz                   & $209$                           \\
    Sampling Rate                   & $\nu_s$                & Hz                    & $16$                            \\
    Integration Time                & $\Delta t$             & s                     & $2.0$                           \\
    Noise Amplitude                 & $\sigma_n$             & Jy                    & $0.65$                          \\
    \begin{tabular}{@{}c@{}}Calibrator Spectral \\
    Flux  Density\end{tabular}      & $S_\nu^\text{CAL}$     & Jy                    & $1.0$                           \\
    Calibrator Position             & $(\alpha,\delta)$      & (deg,deg)             & $(21.0,10.0)$                   \\
    RFI Spectral Flux               & $P_\nu^\text{RFI}$     & $\mu$W.Hz$^{-1}$      & 5.8                               \\
    Orbit Elevation                 & $h$                    & km                    & 20,200                          \\
    Argument of Perigee             & $\gamma$               & deg                   & 5.0                             \\
    Orbit Inclination               & $\beta$                & deg                   & 55.0                             \\
    \begin{tabular}{@{}c@{}}Right Ascension of \\
    the Ascending Node\end{tabular} & $\tilde{\alpha}$       & deg                   & 21.0                             \\
    Initial Gain Amplitude          & $|G|^{(0)}$            & -                     & $\mathcal{N}(1.0, 0.05^2)$        \\
    Gain Amplitude Drift            & $\dot{|G|}$            & $10^{-5}$.s$^{-1}$    & $\mathcal{N}(0.0, 1.0^2)$        \\
    Initial Gain Phase              & $\varphi_{G}^{(0)}$    & deg                   & $\mathcal{U}[-\pi/2, \pi/2]$  \\
    Gain Phase Drift                & $\dot{\varphi_{G}}$    & $10^{-3}$deg.s$^{-1}$ & $\mathcal{N}(0.0, 1.0^2)$        \\
    Noise distribution              & $\eta_{pq}$            & Jy                    & $\mathcal{CN}(0.0, \sigma_n^2)$  \\
    \hline
    Earth Radius                    & $R_e$                  & km                    & $6,371$                          \\
    Earth Mass                      & $M_e$                  & kg                    & $5.9722\times10^{24}$            \\
    Gravitational Constant          & $G_0$                  & N.m$^2$.kg$^{-2}$     & $6.67408\times10^{-11}$          \\
    Speed of Light                  & $c_0$                  & m.s$^{-1}$            & $2.99792458\times10^8$           \\
    \hline
  \end{tabular}}
  \caption{Table of parameter values and distributions for the data generating model. Values have been chosen to, at best, mirror what we have found from various sources including the \href{https://skaafrica.atlassian.net/wiki/spaces/ESDKB/pages/277315585/MeerKAT+specifications}{MeerKAT Specifications} web page.}
  \label{tab:data_gen_params}
\end{center}
\end{table}

We chose the values in Table \ref{tab:data_gen_params} to \comment{represent telescope performance comparable to the real world.} We found values for these parameters from a number of sources. On the \href{https://skaafrica.atlassian.net/wiki/spaces/ESDKB/pages/277315585/MeerKAT+specifications}{MeerKAT Specifications} web page, gain amplitude stability was found to be $<3$\% over 3 hours resulting in $\approx 3\times10^{-6} \text{s}^{-1}$ we use $10^{-5} \text{s}^{-1}$. On the \href{https://ragavi.readthedocs.io/en/latest/gains.html}{\texttt{RAGAVI}} \citep{andati2022ragavi} package web page we found gain amplitudes across antennas to be within 5\% where we have used this value as our standard deviation. On the same web page we found the gain phases between antennas to lie within a 40$^\circ$ band about 0$^\circ$ and we have used 180$^\circ$. The gain phase stability was estimated form the MeerKAT examples on the \href{https://ragavi.readthedocs.io/en/latest/gains.html}{\texttt{RAGAVI}}\citep{andati2022ragavi} web page to be less than 10$^\circ$ over 2 hours resulting in $1.4\times10^{-3}$ deg.s$^{-1}$. We have used $10^{-3}$ deg.s$^{-1}$ as the standard deviation of the gain phase drift. The MeerKAT dish diameter is taken from \cite{jonas2016meerkat}. The observation frequency is chosen to be in the middle of the MeerKAT L-band corrupted by most GNSS signals as is shown in Figure \ref{fig:rfi_lband}. The channel bandwidth is taken from standard L-band 4k mode for MeerKAT. \comment{The visibility sampling rate of 16 Hz is chosen to exceed the minimum requirements (13.7 Hz) for our particular simulation, from Equation \eqref{eq:RFI_sampling_rate}, as calculated in Section \ref{sec:rfi_model}.} The integration time was chosen to be 2 seconds which is one of the options provided by SARAO. The noise is calculated from the estimated SEFD as shown in Section \ref{sec:noise}. The calibrator flux density was chosen to be on the weaker end of the L-band calibrators that SARAO provides on their \href{https://skaafrica.atlassian.net/wiki/spaces/ESDKB/pages/1452146701/L-band+gain+calibrators}{MeerKAT Service Desk} web page. The calibrator position was chosen for convenience in finding a suitable satellite orbit passing within 10$^\circ$. Finally, the RFI orbit parameters are chosen to align with what is publicly available from example TLEs with the argument of perigee and right ascension of the ascending node tuned so that the satellite passes within 10$^\circ$ of both the calibrator and target fields.

\subsection{Bayesian Inference}\label{sec:bayes}

Bayesian inference is a paradigm of statistical inference which uses Bayes' theorem, Equation \eqref{eq:bayes}, 

\begin{equation}\label{eq:bayes}
  \mathcal{P}(\Theta | \mathcal{D},\mathcal{M}) = \frac{\mathcal{L}(\Theta|\mathcal{D},\mathcal{M}) \Pi(\Theta,\mathcal{M})}{Z(\mathcal{D},\mathcal{M})},
\end{equation}
to update our knowledge about some parameters/hypothesis given new information. The goal of Bayesian inference is to acquire the posterior probability distribution, $\mathcal{P}(\Theta|\mathcal{D},\mathcal{M})$, of our model parameters, $\Theta$, given some data, $\mathcal{D}$, and the model, $\mathcal{M}$. The posterior distribution is comprised of three components, the likelihood, $\mathcal{L}(\Theta|\mathcal{D},\mathcal{M})$, the prior, $\Pi(\Theta,\mathcal{M})$, and the evidence $Z (\mathcal{D},\mathcal{M})$. The likelihood is the probability of seeing the data given the parameters in our model. The prior distribution encodes any prior information we have about the parameters of our model. The prior is defined by the user and can include information about the parameters from data not included in the likelihood as well as heuristics or physical limitations of the parameters. The evidence, $Z (\mathcal{D},\mathcal{M}) = \int \mathcal{L}(\Theta|\mathcal{D},\mathcal{M}) \Pi(\Theta,\mathcal{M}) d \Theta$, is a normalizing factor but is also used in model selection problems when deciding between two models, $\mathcal{M}_1$ and $\mathcal{M}_2$, by looking at their ratio. An example would be choosing between a model that contains one satellite compared to two satellites.

In Section \ref{sec:prob_model} we carefully construct a prior that guides our model parameters, $\Theta$, toward desirable solutions and provides suitable initial conditions to reliably find maximum a posteriori (MAP) points through optimization.

An important concept when dealing with a multivariate probability distribution is marginalization. We may consider a subset of our model parameters to be so called nuisance parameters. If we integrate the probability distribution over the nuisance parameters we are left with a marginal distribution over our parameters of interest. The marginal distribution in this case, is a distribution over our parameters of interest taking into consideration all possible values of the nuisance parameters simultaneously. Letting $\Theta=(\Theta_I,\Theta_N)$, where $\Theta_I$ are our parameters of interest and $\Theta_N$ are our nuisance parameters, we obtain our marginalized posterior over $\Theta_I$ by Equation \eqref{eq:marginalize}.

\begin{equation}\label{eq:marginalize}
\mathcal{P}(\Theta_I | \mathcal{D}, \mathcal{M}) = \int
\mathcal{P}(\Theta_I, \Theta_N | \mathcal{D}, \mathcal{M}) \mathop{d\Theta_N}
\end{equation}

\subsection{Likelihood}\label{sec:likelihood}

The likelihood is determined by a combination of our noise model, described in Section \ref{sec:noise}, our forward model and the observed data. Since visibilities have additive noise that is independent and normally distributed in both the real and imaginary parts then our likelihood is the product of the individual likelihood terms for each data point. We further assume that the noise is identically distributed in each data point. The expression for the total likelihood is given in Equation \eqref{eq:likelihood}. 

\begin{equation}\label{eq:likelihood}
    \mathcal{L}(\Theta|V^{obs}) = (\pi\sigma_n^2)^{-N_\mathcal{D}} \exp \Big[
    - \sum_j^{N_\mathcal{D}} \big| V^\text{OBS}_j-\tilde{V}_j(\Theta) \big|^2 /\sigma_n^2 \Big]
\end{equation}

Here we use $\Theta$ to denote the column vector of model parameters, $V^\text{OBS}$ for the observed visibilities, $\tilde{V}$ for the noiseless model visibilities, $j$ to index each data point, and $\sigma_n$ for the standard deviation of the additive complex noise. Our total number of complex-valued data points is $N_\mathcal{D} = N_t N_a(N_a-1)/2$, $N_t$ is the number of time steps and $N_a$ is the number of antennas in the telescope array. For our problem, we only have one frequency channel and polarization.

Each likelihood term is a circularly-symmetric complex normal distribution. Its functional form differs slightly from a (real-valued) normal distribution in that factors of 2 are missing. Since the real and imaginary components are independent with identical variance (circularly symmetric) the variance of the complex visibility sample is twice that of the individual real or imaginary component. Formulating the likelihood in terms of real and imaginary components individually would lead to the factors of 2 returning to the functional form. We therefore consider one complex visibility as one data point as opposed to two, as would be the case for separating the real and imaginary components. Both formulations are equivalent when using the appropriate noise variance, $\sigma_n^2$.

\subsection{Probabilistic Model}\label{sec:prob_model}

\begin{figure*}
  \centering
  \includegraphics[width=0.95\linewidth]{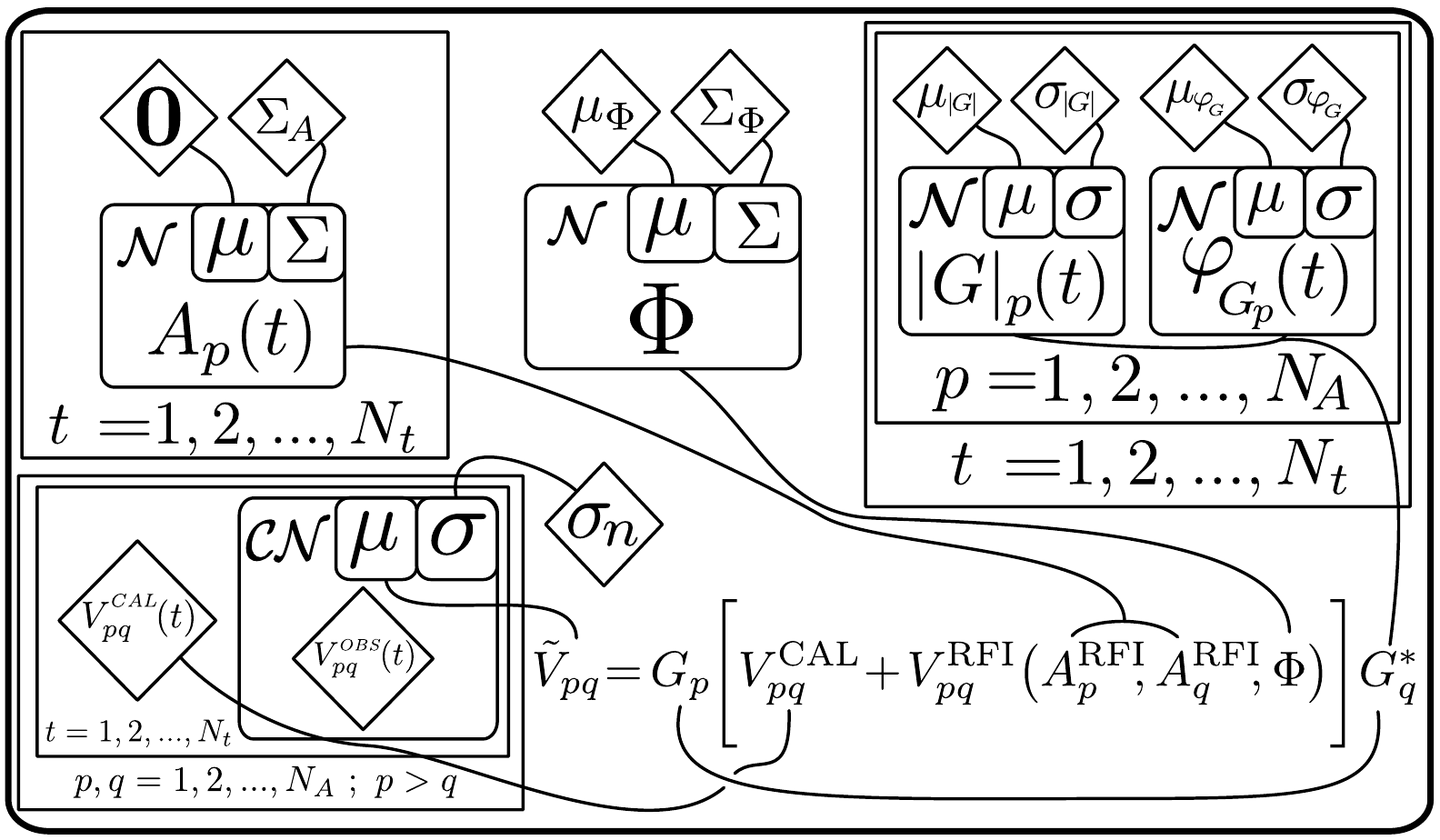}
  \caption{A Bayesian factor graph of the probabilistic model used to estimate uncalibrated and RFI contaminated visibilities. The constants are shown as diamonds. The free parameters of the model are shown as rectangles with rounded corners. Their distributions are shown in the top left corner with the parameterization of the distribution given in the smaller rounded rectangles in the top right corner. Repeated parameters that are indexed are placed in a rectangle, with sharp corners, known as a plate with the index repetition indicated at the bottom centre of the rectangle. The rectangles in the top half of the diagram form the prior over the parameters and the rectangle in the lower left is the likelihood term. The equation in the lower right of the diagram is the mathematical model for the observed data.}
  \label{fig:prob_model}
\end{figure*}

In Section \ref{sec:likelihood} we have described the likelihood. We will now discuss all of the parameters of the forward model and the priors we set on them. Figure \ref{fig:prob_model} shows a Bayesian factor graph (model diagram) that summarizes the entire probabilistic model. This problem is fully constrained and does permit the usage of a purely likelihood based approach or the use of wide, uniform priors. We make use of semi-informative priors based on real-world assumptions that can be made. This improves the consistency and stability of convergence in finding a solution both for optimization and MCMC by regularizing the problem.

% \FloatBarrier

\subsubsection{Gains}

Our gain parameters are composed of amplitudes and phases per antenna, per time step. We have a parameter for the gain amplitude on each antenna at each time step. We have the same for the gain phases except we exclude phases for the last antenna using it as a reference antenna. We set these to 0 in the data generation portion as well as in the forward model. This must be done as our observed visibilities are composed only of differences in phases. A transformation in all gain phases of $\varphi_p' = \varphi_p + \varphi_0$, where $\varphi_p$ is the gain phase on antenna $p$ and $\varphi_0$ is a constant, would leave the measurements unchanged. When $N_a$ is the number of antennas and $N_t$ is the number of time steps, we have $N_t(2N_a-1)$ real-valued parameters to fully describe the complex gains. By using a complex gain parameter for each time step we can model gain variations on shorter time scales than expected thereby not assuming any specific gain variation beyond stability over the individual time step integration time. One could assume the gains to be stable/constant over a 10 second data portion for example and reduce the number of gain parameters by a factor of 10, assuming a 2 second integration time as we use in this paper.

The priors we have on the gain parameters assume reasonable estimates have been made on uncontaminated nearby channels such that the bandpass can be interpolated and provide an estimate in our contaminated channel. \comment{From calibration reports available on the \href{https://archive.sarao.ac.za/}{MeerKAT Data Archive}, most commonly under the observer name `Operator'\footnote{Example observation with a calibration report: \href{https://archive.sarao.ac.za/s3/1622934371-MeerKATReductionProduct/Calibration+Report-1/calreport1.html}{1622934371} (Accessed: 30/04/2023)}, the bandpass variability for the MeerKAT L-band is at most 0.2\%MHz$^{-1}$ and 0.3$^\circ$MHz$^{-1}$ in the gain amplitudes and phases respectively.} We therefore sample from a normal distribution centred on the true value with a 10\% and $10^\circ$ standard deviation for each antenna and use this sample as the mean of our prior distribution for all time steps. The prior standard deviation is set to 10\% of this value for the gain amplitudes and $10^\circ$ for the gain phases. The prior on each parameter is independent and does not include any correlations.

\subsubsection{Satellites}\label{sec:RFI_recovery_model}

To model our satellite-based RFI we have parameters that govern its, per antenna, signal amplitude and parameters that control its orbit around the Earth. For the satellite orbit, we have four parameters as described in Section \ref{sec:rfi_model}. These four parameters describe a circular orbit and are a subset of the six orbital elements needed for a general orbit in the two-body problem. TLEs\citep{TLEs1} expand on these parameters to account for atmospheric drag and the gravitational pull of the moon etc. \href{space-track.org}{Space Track} provides a standard catalog of satellites and their TLEs at constantly updated measurement epochs. In \cite{TLE_error}, over 11k objects from the TLE catalogue are analysed to find their positional uncertainties over a 48 hour period centred on the TLE epoch. These objects have been categorized according to certain orbit characteristics and the standard deviations in their orbit determination have been summarized per class. The standard deviations are quoted in radial, in-track and cross-track (RIC) directions. The RIC coordinates of an orbit, $(\vec{r}_\text{RIC})$, are defined with respect to a reference orbit $(\vec{r}_0)$. The orthogonal RIC coordinate unit vectors are defined in Equation \eqref{eq:RIC_def} and form the rows of the transformation matrix from an Earth-centred inertial (ECI) reference frame, as is used in this paper, to the RIC frame.

\begin{subequations}\label{eq:RIC_def}
    \begin{align}
        \hat{R} &= \vec{r}_0 / |\vec{r}_0| \\
        \hat{I} &= \hat{C} \times \hat{R} \\
        \hat{C} &= (\hat{R} \times \dot{\vec{r}}_0) /|\dot{\vec{r}}_0|
    \end{align}
\end{subequations}

\begin{equation}\label{eq:RIC_frame}
  \vec{r}_\text{RIC} = \begin{bmatrix}
                  \hat{R} \\ \hat{I} \\ \hat{C}
                  \end{bmatrix}(\vec{r}-\vec{r}_0)
\end{equation}

Unfortunately TLEs do not include covariance estimates on their parameters so we make use of those provided in the RIC frame by \cite{TLE_error}. To transform the covariance of an orbit in the RIC frame back to the orbit parameters we make use of the standard error propagation formula assuming normal errors with a minor deviation. Let $\Sigma_\text{RIC}$ be the covariance of a given orbit in the RIC frame. $\Sigma_\text{RIC}$ is a 3x3 matrix and is diagonal for all work in this paper. We wish to transform this into $\Sigma_\Phi$, a 4x4 matrix, which is the prior covariance of our RFI orbit parameters. Firstly, we define our reference orbit $\vec{r}_0(t)$ using the true orbit parameters, given by the TLE in a real-world example. Let $\vec{r}_\text{RIC}(t_j) = T_j(\Phi ; \vec{r}_0(t_j))$ such that $T$ is the function that accepts, $\Phi$, the vector of orbit parameters and produces 3D positions over time in the RIC frame, given in Equation \eqref{eq:RIC_frame}.

We evaluate this at each of the time steps in the calibration data portion. Given each time step has a covariance given by $\Sigma_\text{RIC}$, assumed to be the same at all time steps, we take the average of the transformed precision matrices. The precision matrix is the inverse of the covariance matrix. This leads to a $\Sigma_\Phi$ that when transformed back to the RIC frame at best reproduces the original $\Sigma_\text{RIC}$ but usually has worse constraints in the $\hat{I}$ and $\hat{C}$ directions by a factor of 1.12 and 2.16 respectively. We must do this as using only one time step would not produce a suitable constraint in a higher dimensional orbit parameter space. Equation \eqref{eq:orbit_fisher} shows the formula to generate our prior covariance on the orbit parameters using the method just described.

\begin{equation}\label{eq:orbit_fisher}
  \big( \Sigma_\Phi \big)_{pq} = \left( \frac{1}{n} \sum_j^n \frac{\partial T_j}{\partial \Phi_p} \left(\Sigma_\text{RIC}^{-1} \right)_j \frac{\partial T_j}{\partial \Phi_q} \right)^{-1}
\end{equation}

\begin{figure}
  \centering
  \includegraphics[width=0.45\textwidth]{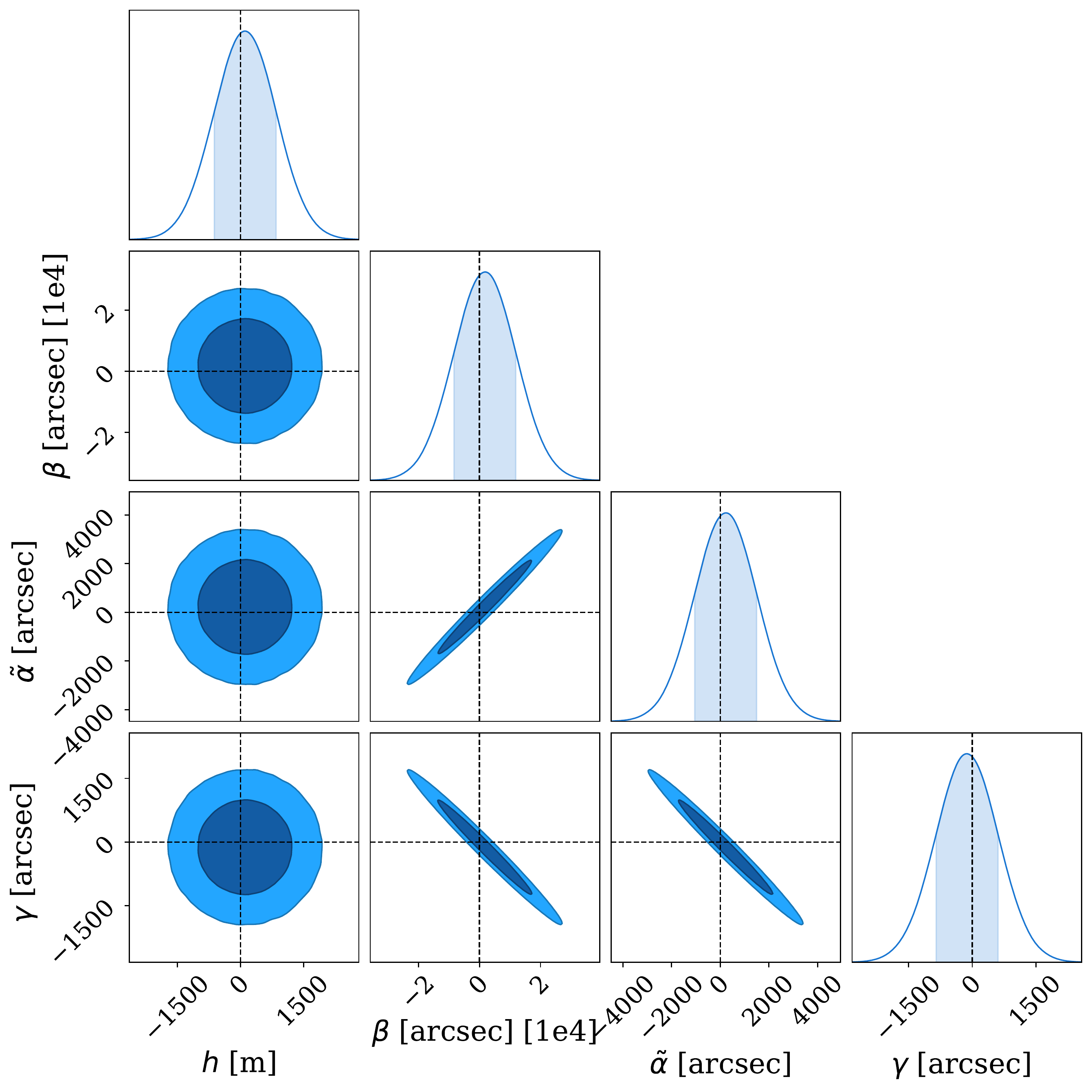}
  \caption{Corner plot showing the prior distribution of the RFI orbit parameters. The strong parameter correlations should be noted as these are crucial in choosing initial parameter values for both the optimization and MCMC routines. The correlations are induced in the error propagation from the co-moving frame to the orbit model parameters. The correlations change depending on the time of the observation. These are for the 0-10 second calibration data portion. The dark and light blue shaded regions show the 68\% and 95\% prior credible regions. The true value and distribution has been shifted to 0 to make the contour levels more readable. }
  \label{fig:cornerPrior}
\end{figure}

In Figure \ref{fig:cornerPrior} we show the prior distribution used for the RFI orbit parameters used in the 0-20s data portion. The covariance provided in \cite{TLE_error} for a medium Earth orbit, as is the case for GNSS satellites, is $\Sigma_\text{RIC} = \bigg(\text{diag}(73, 131, 54)$ m$\bigg)^2$. We chose the prior standard deviations to be 10 times larger than this. We did this to show that our method can handle larger errors in our a prior knowledge than what is publicly available.

% \FloatBarrier

The signal amplitude of the RFI has a parameter per antenna, per time step. We do this as we are modelling the RFI signal amplitude modulated by the primary beam of each antenna, as defined in Equation \eqref{eq:mod_rfi_amp}. 

\begin{equation}\label{eq:mod_rfi_amp}
  A^\text{RFI}_p(t) = E_p \big( \theta(t) \big)\sqrt{I_p^\text{RFI}(t)}
\end{equation}

This is a more general approach as compared to assuming something about the primary beam of the antenna/s and/or the intrinsic RFI signal. Modelling each of these separately would be degenerate as we can only constrain the product of the beam and the RFI signal. If, in our subtraction, we were to assume the primary beam is the same across all antennas we could parameterize the primary beam using a Zernike polynomial based model as in \cite{zernikebeam}.

For our prior on the modulated RFI amplitudes we choose an uninformative prior, Equation \eqref{eq:rfi_amp_prob}, with a very wide range. 
\begin{equation}\label{eq:rfi_amp_prob}
    A_p^\text{RFI}(t_j) \sim \mathcal{N}(0,\Sigma_A)
\end{equation}
We chose each parameter prior to be a normal distribution with mean 0 $\sqrt{\text{Jy}}$ and covariance, $\Sigma_A$, to be fully independent with diagonal values of 10 000 Jy. Each parameter prior is fully independent. \comment{Therefore, no prior information about the primary beam or the RFI signal is assumed, even the choice of 10 000 Jy as the variance is somewhat arbitrary and chosen to be large enough to not influence our subtraction results.} It is possible to place a correlated (in time or across antennas) prior on these parameters that leads to better posterior constraints on the gain amplitude parameters at the expense of introducing slight biases in the results as this correlation assumptions is not always true.

\comment{We have not modelled the $A_p^\text{RFI}(t_j)$ parameters as complex values in our subtraction analysis. This, however, would be a requirement when applied to real data due to the complex nature of the primary beam. We leave this extension for future studies and is expected to be possible through the use of suitable priors on the $A_p^\text{RFI}(t_j)$ parameters. Furthermore, an additional complexity to be considered in future studies are the cross-polarization terms in the primary beams. The inclusion of these effects in the forward model is likely to render the use of the RFI signal for calibration purposes invalid. However, the subtraction of the RFI signal will still be possible and is in actuality the most important aspect of this method.}

A very important aspect of our RFI model is to model the time-smearing that occurs due to the rapidly varying phases induced by the fast moving RFI. We, therefore, evaluate the position of the satellite at multiple time points centred about each time step in our data. Additionally, we perform a linear interpolation, and extrapolation at the edges of the data portion, for our RFI amplitudes. This is needed due to the rapidly varying modulated amplitude caused by, at minimum, the movement of the RFI source through the primary beam. Once the amplitudes have been re-sampled at the higher rate, equal to the position sampling, RFI visibilities are predicted at this finer rate and then averaged back down to the cadence of the observed visibilities.

\subsection{Optimization and Laplace Approximation}\label{sec:optimize}

We developed our forward and probabilistic models using the \JAX\ python library \citep{JAX,JAXgithub} so that we could make use of just-in-time (JIT) compilation and automatic differentiation. This allowed us to speed up our computation dramatically and get exact derivatives for our own optimization routine. In this section we will give a brief explanation of the Laplace approximation and the custom optimization routine that we developed for this work.

For a quick approximation of the posterior distribution we can make use of the Laplace approximation \citep{tierney1986laplace}. The Laplace approximation approximates the posterior distribution as a multivariate normal distribution centred on the maximum a posteriori (MAP) point, $\Theta_\text{MAP}$. Since a normal distribution is defined by its first and second moments we must find these. We therefore make use of an optimization scheme to find the MAP position (equivalent to the mean, first moment, of a normal distribution). At this point the covariance of a multivariate normal, its second moment, is equivalent to the negative inverse of the Hessian of the log posterior. Symbolically this is described in Equation \eqref{eq:laplace} where $p(\Theta | \mathcal{D})$ is the posterior distribution density function:

\begin{equation}\label{eq:laplace}
  p(\Theta | \mathcal{D}) \simeq \mathcal{N} \left( \Theta; \Theta_\text{MAP}, \Sigma_\text{MAP} \right), \quad \Sigma_\text{MAP}^{-1} = -H \left(\ln p \left(\Theta | \mathcal{D} \right) \right)
\end{equation}

In Equation \eqref{eq:laplace} we make use of the Hessian, $H$, which is the matrix of second order partial derivatives of a scalar valued function, $f$, as defined in \eqref{eq:hessian} where $j$ and $k$ are the row and column indices of the matrix and $\Theta_j$ is the $j^\text{th}$ parameter in the vector of parameters $\Theta$.

\begin{equation}\label{eq:hessian}
  H_{jk}(f) = \frac{\partial^2 f}{\partial \Theta_j \partial \Theta_k}
\end{equation}

Therefore, given the MAP position by running an optimization scheme using the negative log posterior (NLP) as the minimization surface and then evaluating the Hessian of this surface at the MAP we can obtain the Laplace approximation to our posterior distribution. The Laplace approximation is exact when the posterior distribution is exactly Normal. For all but the simplest problems this is not the case, however, in the limit of infinite data with finite variance the posterior distribution tends toward a normal distribution thanks to the central limit theorem. Next we describe the optimization routine used to find the MAP point. In Section \ref{sec:MCMC_chains}, we show that this is in fact an excellent approximation of our posterior by comparing the Laplace approximation to the full posterior obtained using a MCMC approach.

There are many optimization routines publicly available, however, many of these did not perform particularly well on our  problem out of the box. Given that we are able to evaluate the Hessian exactly we developed our own quasi-Newton method. Our method evaluates the Hessian periodically to save on computation. The update step for a general quasi-Newton scheme is
\begin{equation}\label{eq:quasi-newton}
  \Theta_{k+1} = \Theta_k - \epsilon B^{-1}\nabla f(\Theta_k) \, ,
\end{equation}
where $f$ is the function to minimize, $B$ is the Hessian approximation, $\Theta_k$ is the parameter vector at step $k$, and $\epsilon$ is the step size. Since inverting the Hessian becomes very expensive in a high dimensional parameter space, and will not always produce a suitable $B^{-1}$ in problematic sections of the parameters space, we block diagonalize the Hessian before inverting it. This allows us to invert smaller sub matrices, the blocks, and then recompose them to make a full $B^{-1}$. We do this for the gain amplitudes, phases, RFI amplitudes, RFI orbit parameters blocks separately. This reduces the computational expense but more importantly leads to a more robust optimizer while still efficiently navigating the parameter space. When we reach a part of the parameter space that is better behaved,  we use the full Hessian and then invert it using an eigendecomposition and applying the \texttt{softabs} \citep{betancourt2013general} function to the eigenvalues before taking their reciprocals. This ensures that we have a $B^{-1}$ that is positive semi-definite. This is usually when $\chi^2_\text{dof}<1.1$. The final stages of optimization using the full Hessian inverse allow us to more efficiently converge on the MAP position.

We calculate $B^{-1}$ every 250 update steps and find that 500 steps using the block diagonal version is enough, after which less than 250 steps are typically needed using the full inverse. We use a decaying step size $\epsilon$ that is reduced by an order of magnitude when using the full inverse in the final optimization steps. Using this scheme we were able to achieve excellent convergence with $\chi^2_\text{dof} \approx 1.01$ in less than 1 minute per 10s data portion using a laptop with 16GB of RAM. The covariance estimation using the Hessian takes around 10 seconds per data portion of $\approx$ 1000 parameters.

Optimization routines are typically very sensitive to the initial parameter values and ours is no exception. We sample 10k points from a region centred about the true parameter values, for the gains and RFI orbit parameters, with standard deviations one quarter of those used for their respective prior distributions. This should not be a problem when applied to real data as we expect to know these parameter values to this accuracy or better, we just use especially wide priors in our analysis. The initial RFI amplitude parameters are calculated at each time step using the data and the initial gain parameters, and are the same across antennas. We evaluate the NLP for each of the 10k parameter sets and start from the best position and run the optimizer till convergence. Convergence is assumed when, $\chi^2_\text{dof} < 1.05$, the improvement in the NLP is less than a set threshold, and the Hessian at that location is positive definite. If the optimizer does not converge according this criteria the next best initial position is used to run a new optimization round.

\subsection{MCMC Implementation}\label{sec:MCMC}

In this section we describe the Markov Chain Monte Carlo (MCMC) implementation we have used in our analysis. We explain the advantages and highlight some of the specifics of how we obtained our results that are given in Section \ref{sec:MCMC_chains} where we also analyze its accuracy and convergence for our problem. In this paper we have made use of Hamiltonian/Hybrid Monte Carlo (HMC) for sampling the posterior. We designed our own HMC implementation using the \JAX\ python library \citep{JAX,JAXgithub} so that we could customize it as we saw fit as well as make use of just-in-time (JIT) compilation and automatic differentiation.

The benefit of HMC over a Metropolis-Hastings algorithm using random walk is that successive proposals in HMC are distant from one another, significantly reducing their correlation and leading to higher effective sample sizes for a given MCMC chain length. HMC is a Monte Carlo method where sample proposals are generated by treating the parameters as position coordinates of a particle and the negative log posterior (NLP) as a potential energy function, $U(\vec{x})$. New proposals are generated by sampling associated momenta from a predefined distribution where its negative log represents the kinetic energy of the particle. A proposal is then formed by evolving the particle's position using Hamiltonian dynamics. Equations \eqref{eq:hamilton} show Hamilton's equations 

\begin{subequations}
  \label{eq:hamilton}
  \begin{align}
    \frac{d\vec{x}}{dt} &= \frac{\partial \mathcal{H}}{\partial \vec{p}} \label{eq:postion_diff} \\
    \frac{d\vec{p}}{dt} &= -\frac{\partial \mathcal{H}}{\partial \vec{x}} \label{eq:momenta_diff}
  \end{align}
\end{subequations}

where $\vec{x}$ is the position in parameter space, $\vec{p}$ is the momentum and $\mathcal{H}$ is called the Hamiltonian which is defined as the sum of the potential energy and kinetic energy, Equation \eqref{eq:hamiltonian}. 

\begin{equation}\label{eq:hamiltonian}
  \mathcal{H} = U(\vec{x}) + \frac{1}{2} \vec{p}M^{-1}\vec{p}
\end{equation}

The momentum is defined as $\vec{p}=M\cdot d\vec{x}/dt$ where $M$ is the mass matrix. The samples generated from HMC have position and momenta. We can then marginalize over the momentum variables leaving our position variables which are our parameters` samples.

For all but the simplest of posterior distributions Hamilton's equations must be numerically integrated to evolve the position and momentum variables. A symplectic integrator with time reversibility is needed for this \citep{neal2011mcmc}. Since our Hamiltonian is separable we have used the leapfrog integration scheme. By using a numerical integration scheme an error is introduced that is dependent on the integration step size. Due to this a Metropolis-Hastings acceptance test must be introduced. The acceptance probability, $\alpha$, of a proposal is given by Equation \eqref{eq:acpt_prob}:
\begin{equation}\label{eq:acpt_prob}
  \alpha = \min \Bigg(1, \frac{\exp(\mathcal{H}_f)}{\exp(\mathcal{H}_i)} \Bigg) \, ,
\end{equation}
where $\mathcal{H}_i$ and $\mathcal{H}_f$ are the initial and final values of the Hamiltonian in a single proposal evolution. In high dimensions the optimal acceptance rate for HMC is $\approx 65\%$ \citep{beskos2013optimal}.

For our particular HMC implementation we have used the standard kinetic energy function with a mass matrix including off-diagonal terms. This leads to sampling momenta from a multivariate normal distribution. We define the mass matrix to be independent of position (Euclidean HMC) leading to a simpler implementation that still allows us to take parameter correlations into consideration. In such a formulation, parameter correlations that change over the parameter space cannot be taken into account. As such, the behaviour of the posterior can significantly affect the efficiency of the sampler. Generally, when the information content of the data is high, the parameter space close to the maximum a posteriori (MAP) point is approximately Gaussian and sampling with HMC is very efficient. Finding this portion of the parameter space, close to the MAP, can often be the toughest part of the problem.

More sophisticated HMC routines like Riemannian Manifold Hamiltonian Monte Carlo \citep[RMHMC,][]{RMHMC}, where the mass matrix is a function of position, exist and allow efficient sampling of highly non-Gaussian posteriors. Fortunately, such a routine is not needed for our problem as our posteriors are very well approximated by multivariate normal distributions near the maximum a posteriori (MAP) point, as is shown in Section \ref{sec:MCMC_chains}.

\subsubsection{Initial Conditions}

By initial conditions we mean the initial position and mass matrix. The determination of these is crucial for our problem. When using unsuitable initial positions the HMC struggles to find the typical set\footnote{The typical set of a distribution is the volume of parameter space in which nearly all of the probability mass is located.} reliably. Additionally, even with a suitable initial position, the auto-correlation times of many parameters would be unacceptable for real-world usage/implementation. The solution to this second part is tuning the mass matrix to include information about the parameters scales and correlations. Ideally (in the Euclidean HMC formulation), the mass matrix is chosen to be as close as possible to the posterior inverse covariance matrix.

The most difficult set of parameters to tune the initial conditions for are the satellite orbit parameters. Very small changes in these parameters lead to large differences in the likelihood. Additionally, these parameters are very strongly correlated leading to very inefficient sampling when the chosen mass matrix does not incorporate the correct correlations.

Our initial sampling location is chosen by sampling the gain amplitudes, phases and RFI orbit parameters from the prior. The initial modulated RFI amplitude parameters are estimated by using the data, the other initial parameters, and the calibrator visibilities and performing a rough calculation to get RFI amplitudes per time step. We use this estimate, at each time step, and use the same across all antennas. The resulting estimate is always positive which is not a problem as our likelihood cannot tell the difference between positive and negative RFI amplitudes.

Now that we have a suitable initial position we are left with choosing a suitable mass matrix. We use the block diagonalized inverse Hessian as described in Section \ref{sec:optimize} for the mass matrix. Using this mass matrix we can evolve the HMC sampler until it has found the typical set. To help this process we appropriately vary the integration step size of the HMC sampler. Once the typical set has been found we can re-evaluate the Hessian at our MAP sample point achieved thus far. We continue to vary the integration step size until an optimal value is reached (leading to an acceptance rate of $\approx 65\%$). Once this criteria is achieved everything is in place to start efficiently sampling from the posterior. All samples prior to this point are regarded as burn-in\footnote{Burn-in samples are initial samples in an MCMC chain that are discarded as they would skew our posterior sample estimate. In our case the initial samples do not conform to detailed balance due to the variation of the integration step size and mass matrix.}. Throughout the sampling the number of integration steps are kept fixed. Once the burn-in period is complete the integration step size and the mass matrix are also fixed. This is done to preserve detailed balance from this point on ensuring that our samples converge to the posterior distribution. We do this for multiple chains in parallel with different random seeds to attain robust results.

\section{Results}\label{sec:results}

\comment{The goal of this section is to explore, in detail, the performance of our method (\tabascal), the robustness of our uncertainty estimates and the accuracy of our posterior approximation. To this end, this section is split in two. In the first subsection, we analyze the bias and standard deviations in our parameters of interest, the antenna gains. In the second subsection, we compare our posterior results derived from MCMC and the Laplace approximation. We also show the reliability and accuracy of our Markov chain Monte Carlo (MCMC) posteriors by calculating the uncertainty in the posterior standard deviations as well as the degree of convergence in our MCMC chains.}

Throughout this section we will often reference the bias in a parameter, both relative and absolute \footnote{What we refer to as bias in this text is often referred to as error in other texts.}. For clarity, the bias refers to the difference between the estimated value, $\hat{x}$, and the true value, $x^\text{true}$, i.e. $\hat{x}-x^\text{true}$ and the relative bias is then $(\hat{x}-x^\text{true})/x^\text{true}$. Letting $\hat{\sigma}_x$ be the estimated posterior standard deviation/uncertainty in parameter $x$, we formulate the normalised bias as $(\hat{x}-x^\text{true})/\hat{\sigma}_x$. For a normally distributed, unbiased, estimator with reliable uncertainties we expect the normalised bias to conform to a standard normal distribution. When referring to the standard deviation on a gain amplitude it will always be quoted in \% as an uncertainty relative to the true parameter value.

\subsection{Posterior Results for Calibration}\label{sec:posterior}

In this section we analyze the posterior distributions and discuss how to use their estimates. We show that the estimation biases are consistent with zero (i.e. are within the uncertainty estimates), and that the standard deviations reduce with increasing signal-to-noise ratio (SNR), where the signal is the total observed signal (i.e. the contaminated visibility).

\begin{figure}
  \centering
  \includegraphics[width=0.45\textwidth]{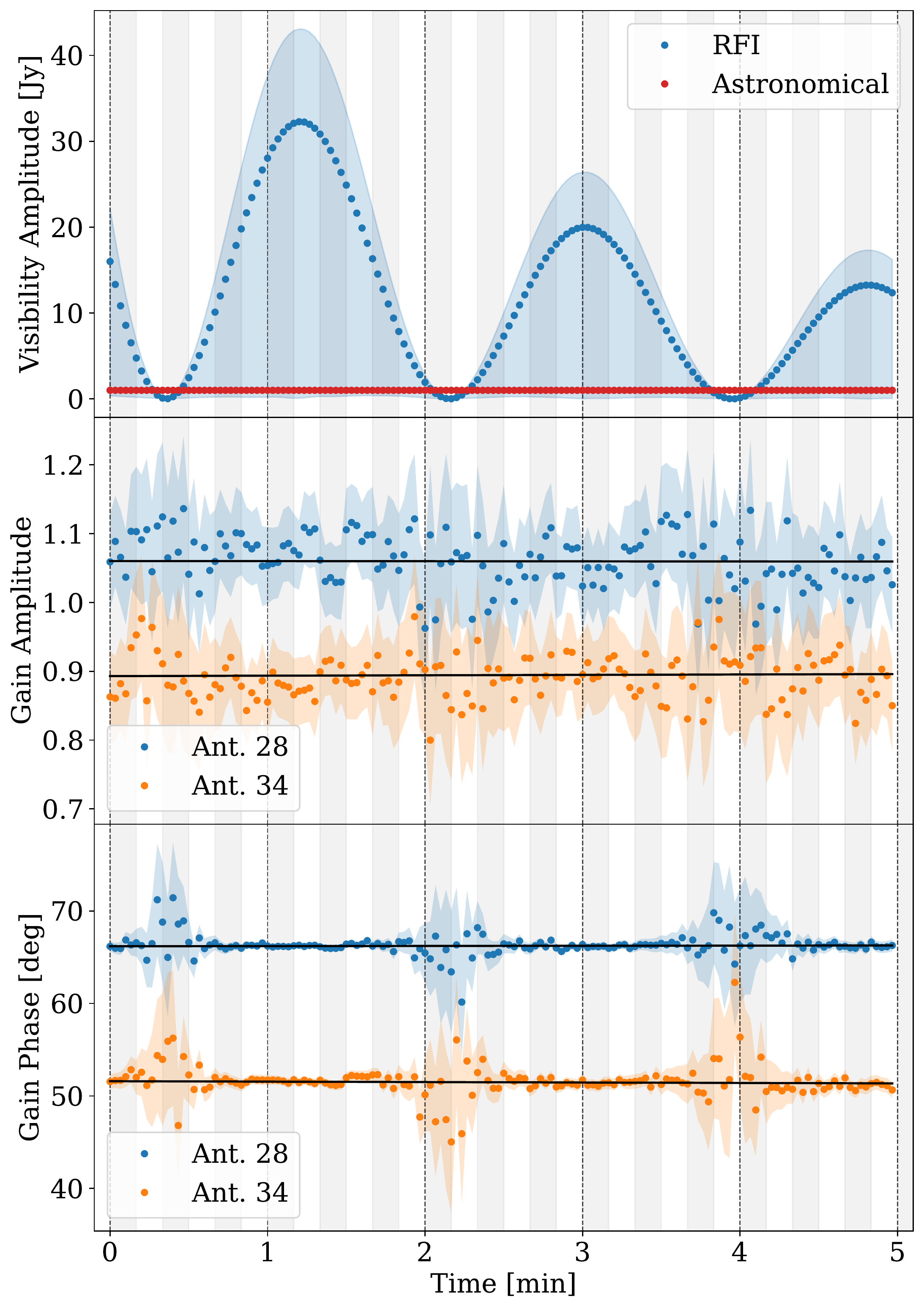}
  \caption{Estimated calibration solutions over time: note how the uncertainties on the gain amplitude and phase are minimised when the RFI is strong. The top panel shows the visibility amplitudes over time in the calibration observation for the RFI and astronomical contributions. The red and blue dots are the average amplitude across all baselines and the light blue shaded region shows the amplitude range across all baselines. The strong amplitude variation of the RFI visibilities is due to the satellite passing through the primary beam sidelobes. The middle and bottom panels show the Laplace estimated gain amplitudes and phases respectively. The light blue/orange shaded error regions in these panels give the $2\sigma$ credible intervals from the posterior distributions. The specific example antennas shown are chosen to create a more legible plot. The vertical interleaved grey and white shaded strips show the 10s data portions used in each optimization run.}
  \label{fig:vis_cal}
\end{figure}
% \FloatBarrier

\begin{figure}
  \centering
  \includegraphics[width=0.45\textwidth]{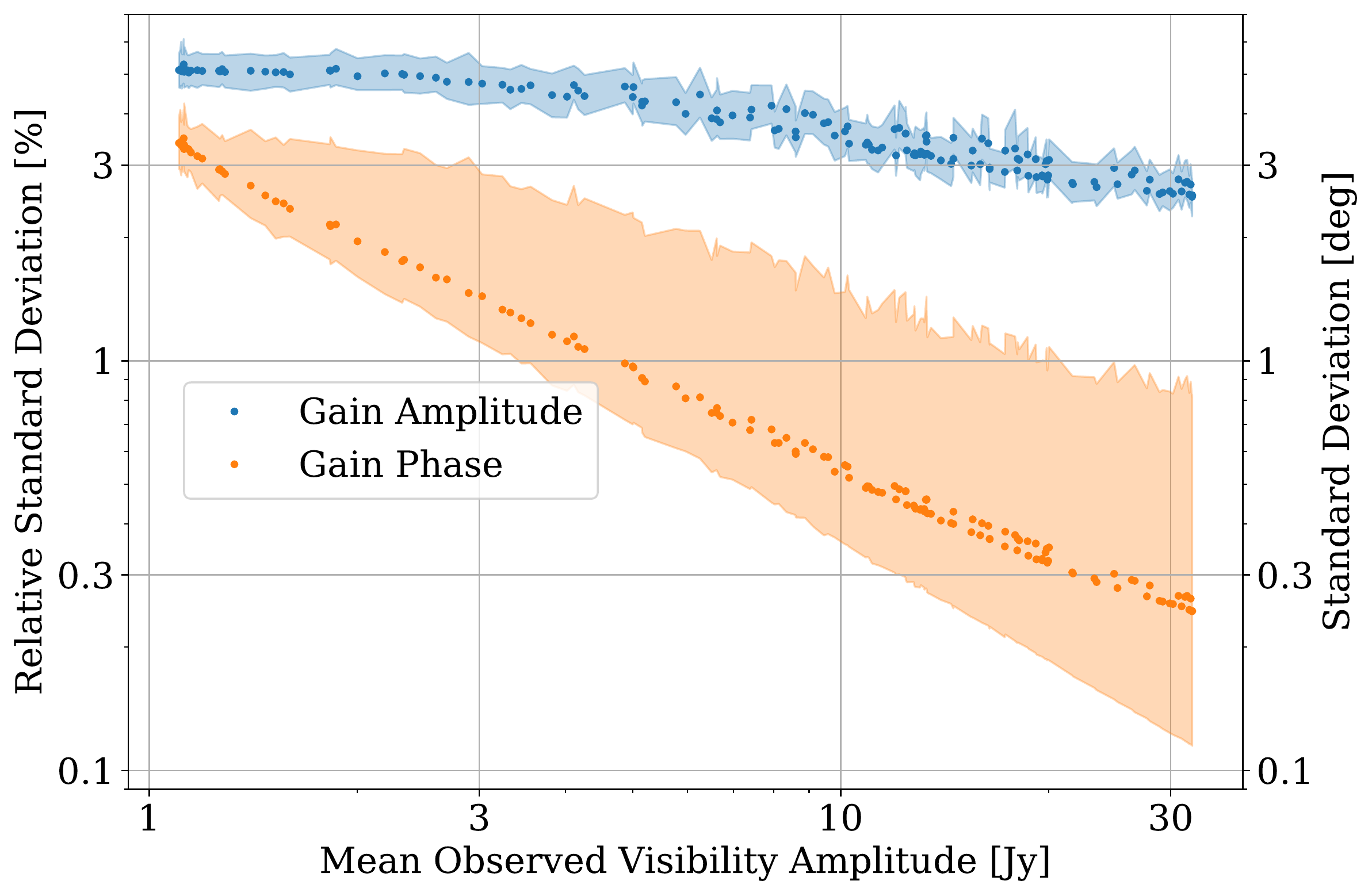}
  \caption{How total signal, including RFI, improves calibration constraints: using a common gain parameter for both the RFI and astronomical contribution (since they are within $10^\circ$ on the sky) allows us to leverage the total signal to improve calibration constraints. The posterior standard deviations in gain parameters are plotted against the mean (over baseline) observed visibility amplitude. The dots show the mean standard deviation across antennas and the shaded region of the corresponding colour show the minimum and maximum range (in uncertainty) across antennas. The parameter standard deviations are per time step with a 2 second integration time where the calibrator flux is 1 Jy and noise level is 0.65 Jy. The prior standard deviations were 10\% and 10$^\circ$ for the amplitudes and phases respectively.}
  \label{fig:SDvsVis}
\end{figure}

Figure \ref{fig:vis_cal} shows a summary of the data and gain estimates over a 5 minute calibration observation for which the simulation details are given in Section \ref{sec:data_gen}. \comment{The main takeaway from Figure \ref{fig:vis_cal} that is highlighted in Figure \ref{fig:SDvsVis} is that the gain estimates are better constrained when the RFI signal is stronger and are bounded from below by the signal strength of the calibrator. We, therefore, always achieve calibration constraints at least as good as an uncontaminated observation. This is caused by both signals sharing the same gain parameters in the model.} The estimates in this figure are obtained by using the Laplace approximation on 10 second portions of the data in parallel. \comment{We use 10 second data portions as this provides enough SNR to constraint the RFI orbit parameters even when passing through a null of the beam. This time should be adapted to the situation where the only downside to increasing the time is reduced parallelization.} Each portion uses the same RFI parameter priors. The gain amplitude and phase priors have the same standard deviations (relative and absolute respectively) across all data portions. The gain prior means are offset by the same, relative and absolute, quantities per antenna in all portions. Each data portion used an individually calculated RFI visibility sampling rate to optimize run times and memory usage according to the required accuracy. \comment{Equation \eqref{eq:RFI_sampling_rate} shows the calculation used to determine the minimum sampling rate where we have used the maximum observed visibility amplitude in lieu of the RFI visibility amplitude assuming the average gain amplitudes are 1. The use of the total visibility amplitude only leads to a higher than necessary sampling rate and also provides a lower bound when the RFI is passing through an antenna null.}

Due to both the minimum RFI sampling rate and inversion of an $N_p \times N_p$ matrix, it is more efficient to perform estimation on portions of data from a memory and computation time stand point. This parallelization scheme poses many benefits including increasing robustness to failures in optimization. Failed optimizations can be rerun with initial positions informed by the successful runs. Additionally, gain solutions can be estimated on the fly instead of waiting for all the data to be available.

A notable observation from Figure \ref{fig:vis_cal} is that the biases and standard deviations for the gain estimates decrease for increasing RFI visibility amplitude. This indicates that the model is able to leverage the added signal from the RFI to increase the SNR. Figure \ref{fig:SDvsVis} shows this relationship clearly. We obtain tighter constraints on the gain parameters with increasing observed visibility amplitude. This shows that we are using the total signal to calibrate the antennas. \comment{The gain phase constraints have a much stronger correlation with the observed signal compared to the gain amplitudes. This is because the gain phases are linked across time due to the common RFI orbit parameters. This is not the case for the gain amplitudes due to the separate RFI amplitude parameters at each time step and antenna.} The prior standard deviations, $10\%$ and $10^\circ$, in the gains are both larger than the posterior uncertainties, at all signal levels. This shows that our constraints are not strongly affected by the priors in the weak-RFI regime. \comment{In Figure \ref{fig:SDvsVis}, the spread in posterior uncertainties, at each visibility amplitude, is due to the spread in SNR for different antennas. The antennas in the core of the array contribute mostly to shorter baselines that have a larger RFI signal compared to longer baselines due to time-smearing of the signal.}

% \FloatBarrier

\begin{figure}
  \centering
  \includegraphics[width=0.45\textwidth]{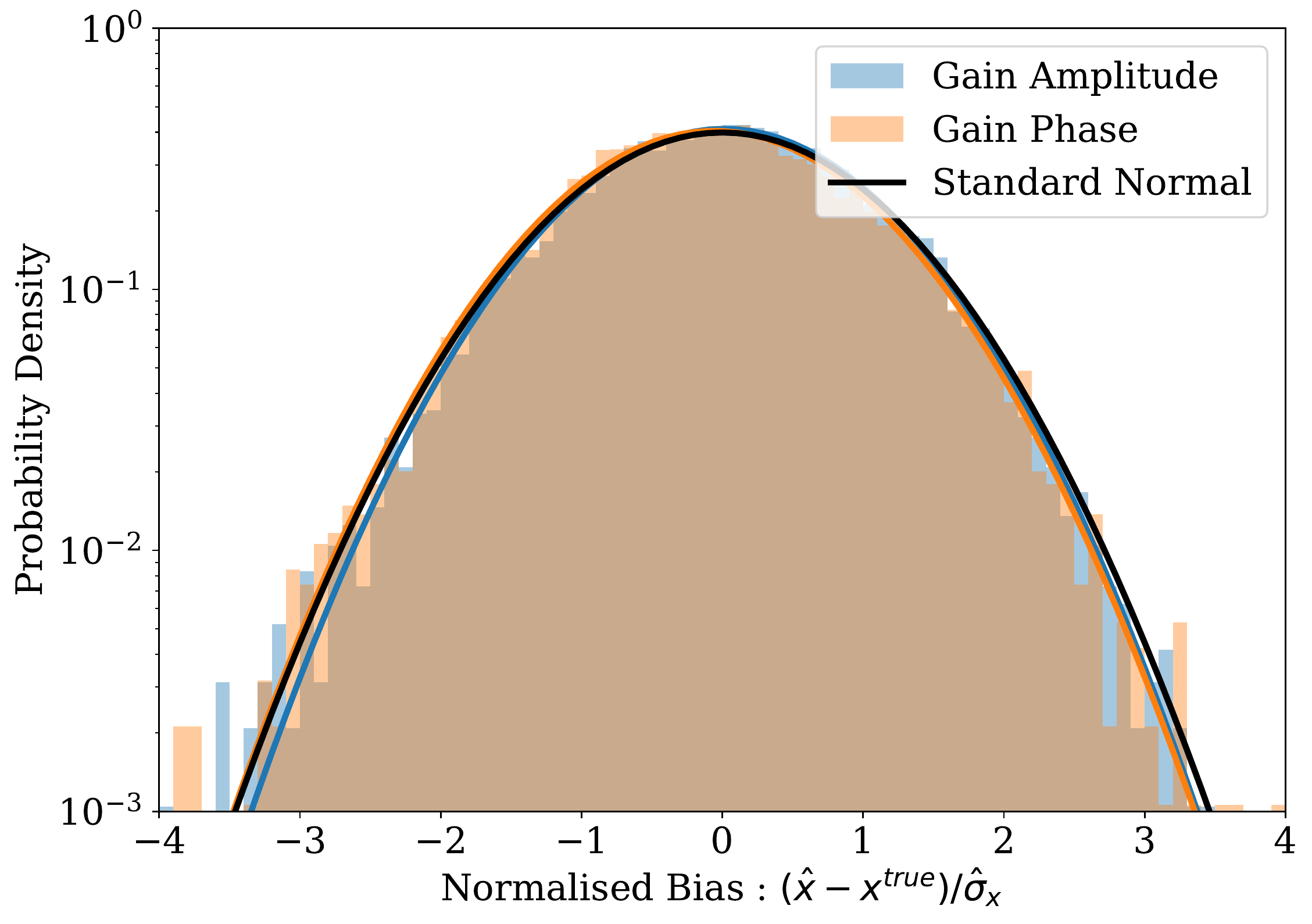}
  \caption{Histograms showing \tabascal\ gain estimates are unbiased and have reliable uncertainties. Distribution in the normalised bias of the estimated gain amplitudes and phases. These are for the 5 minute calibration observation displayed in Figure \ref{fig:vis_cal}. For an unbiased estimation of the parameters we expect the normalised bias to follow a standard normal distribution, which it does to good accuracy. The means and standard deviations of the normalised bias distribution for the gain amplitudes are $0.01 \pm 0.97$ and the gain phases are $-0.06 \pm 0.99$.}
  \label{fig:norm_bias}
\end{figure}

\comment{Figure \ref{fig:norm_bias} shows that the biases and associated standard deviations, on the gain parameters, are statistically consistent. This is shown by the mean and standard deviation of the distributions being consistent with zero and one respectively.} In Section \ref{sec:MCMC_chains} the quality of the Laplace approximation is analyzed and we find the accuracy to be sufficient for practical purposes.

% \FloatBarrier

Typically, the gain estimates from the calibration observation would be averaged per antenna under the assumption they are constant. Section \ref{sec:comb_ind} describes how to combine our posterior estimates from different data portions as these are independent. The combined estimate takes into consideration the correlations between our parameter estimates to maintain reliable uncertainties. Once appropriate data portions are combined, the different time steps within a data portion can be combined by following the recipe outlined in Section \ref{sec:comb_corr}. The second step requires the extraction of the subcovariance matrix of the parameter estimates associated with a single antenna. After this procedure is complete one is left with per antenna gain estimates using the full calibration observation. \comment{We have not shown such combined estimates as our gains are varying over time at a rate such that 0.3\% and 0.3$^\circ$ changes can be expected, in amplitude and phase respectively, over a 5 minute interval. Since the combined estimates predict uncertainties of this order or lower, these variations would bias our combined estimate leading to non-representative uncertainties.}

\begin{figure}
  \centering
  \includegraphics[width=0.45\textwidth]{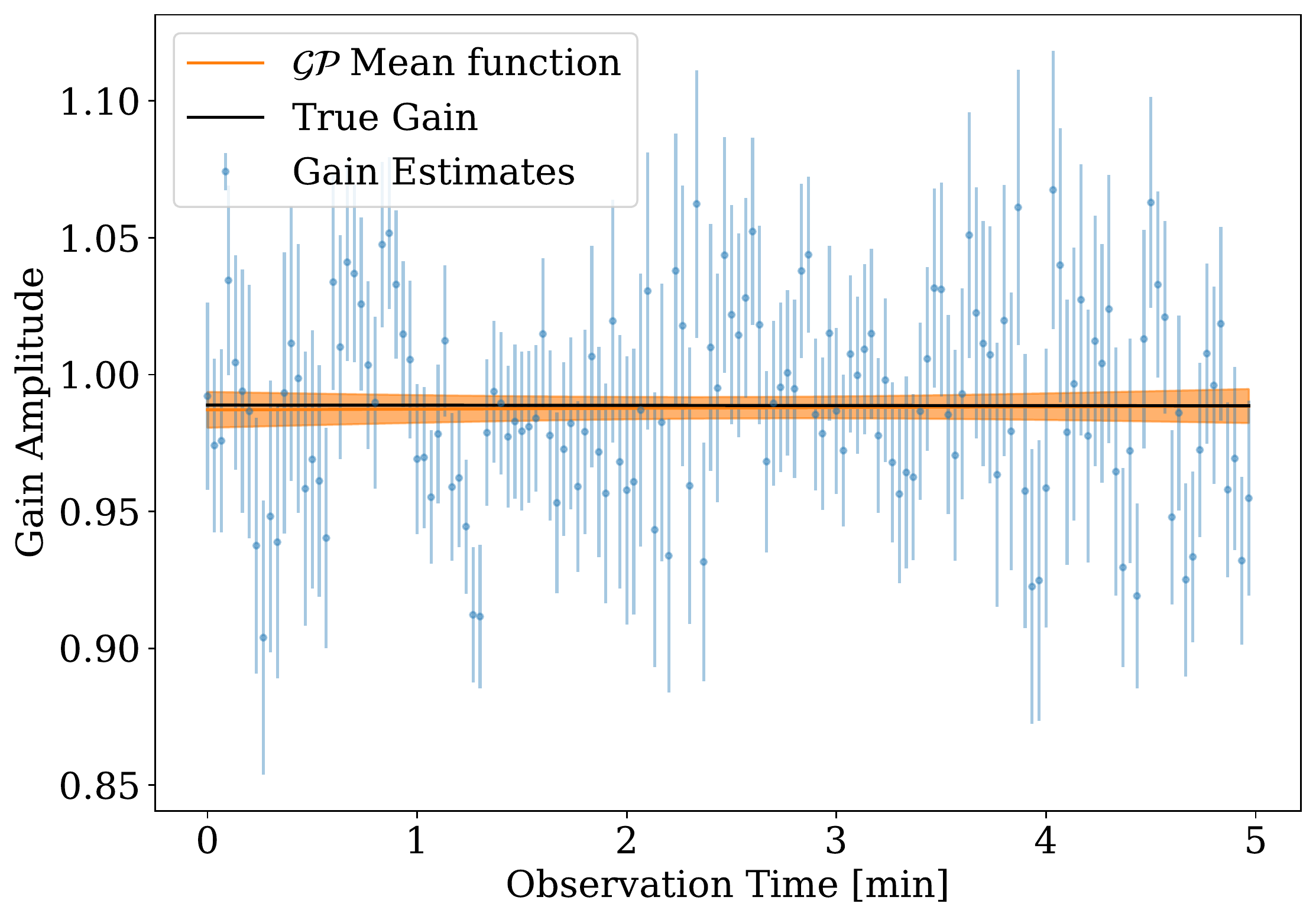}
  \caption{An example of a Gaussian process that has been fitted to the gain amplitude estimates of antenna 22 using the full posterior covariance. The blue dots with error bars are the posterior estimates as shown in Figure \ref{fig:vis_cal}. The orange curve and shaded region is the fitted Gaussian process with its 68\% credible interval. The black line is the true gain phase used in the data generation. The root mean squared error (RMSE) for the Gaussian process is 0.46\%, in the calibration portion, aligning well with its mean uncertainty of 0.45\%. 20 minutes after the calibration observation the uncertainty and RMSE grow to around 2\%.}
  \label{fig:exGP}
\end{figure}

For our case, the preferred procedure, in our opinion, is to fit a Gaussian process (GP) to the estimates and potentially other calibration observation simultaneously. The covariances of each data portion would be used as the noise parameter in the GP. The resulting gain estimates, with covariances, for the sandwiched target observation can be used as an informative prior in the 2nd Generation Calibration\footnote{2nd Generation Calibration is another term for \texttt{selfcal} in the direction independent calibration regime.} (2GC) process. Figure \ref{fig:exGP} shows an example GP for reference where the marginalized posterior subcovariance and mean has been used to fit gain amplitudes in a single (5 min.) calibration portion. We find that fitting a Gaussian process (combining estimates) reduces uncertainties to around 0.5\% and 0.05$^\circ$ for the gain amplitudes and phases respectively. The fitted GP model can be used to predict gains in the near future/past where its uncertainties increase by an order of magnitude 20 minutes away. 

\begin{figure}
  \centering
  \includegraphics[width=0.45\textwidth]{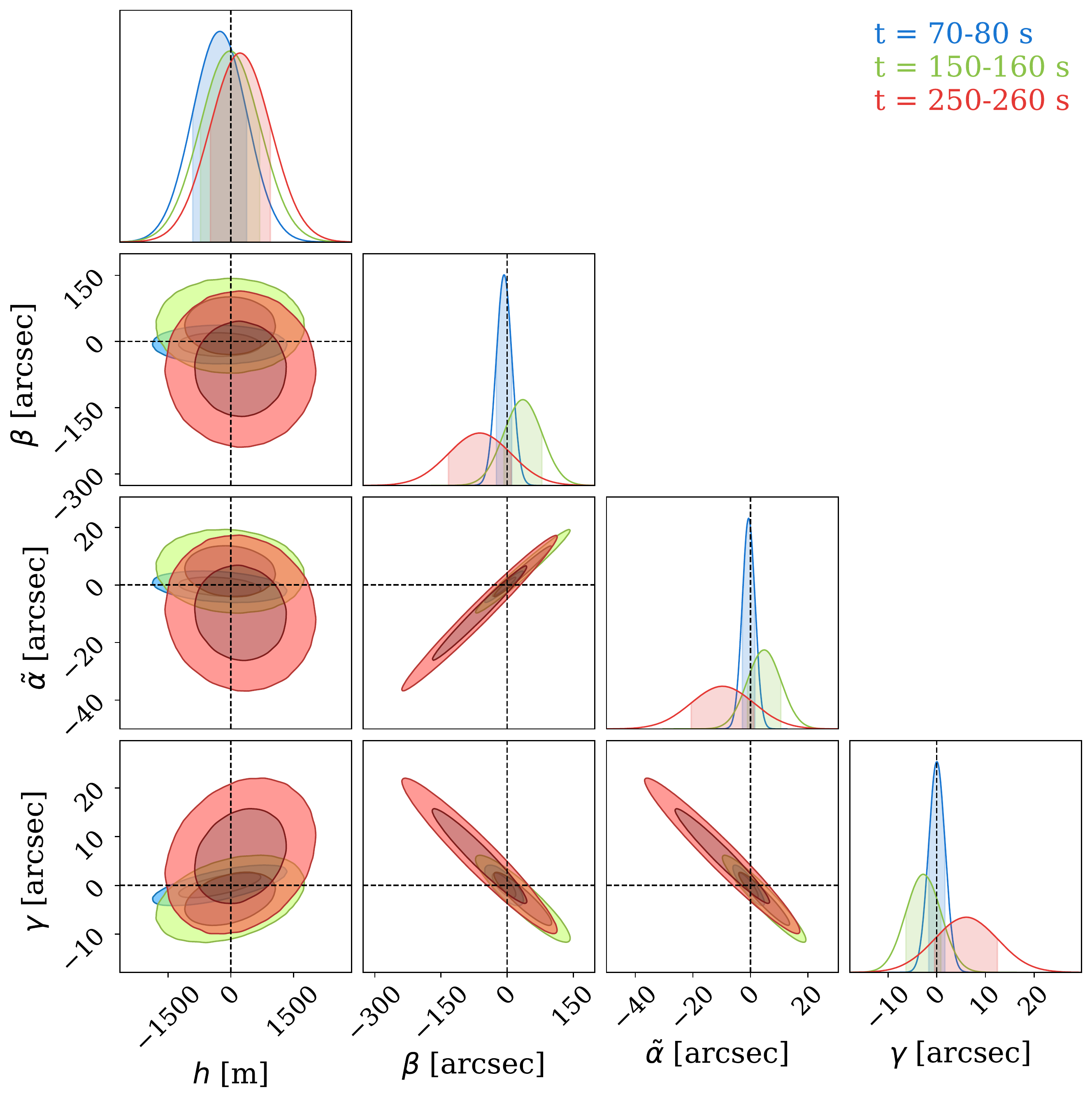}
  \caption{Posterior marginal distributions of the RFI orbit parameters from three different 10s data portions showing the variation in constraints and parameter correlations. Note the constraints on the angular parameters have improved by 2 orders of magnitude compared to the prior in Figure \ref{fig:cornerPrior} and improve a further 2 orders of magnitudes when combined to form Figure \ref{fig:RFIcorner5m}. The times for the data portions are shown in the top right corner. The distributions and true value have been shifted to make the true parameter values 0. This is done to make uncertainties more legible on the axes.}
  \label{fig:RFIcorner3}
\end{figure}

\begin{figure}
  \centering
  \includegraphics[width=0.45\textwidth]{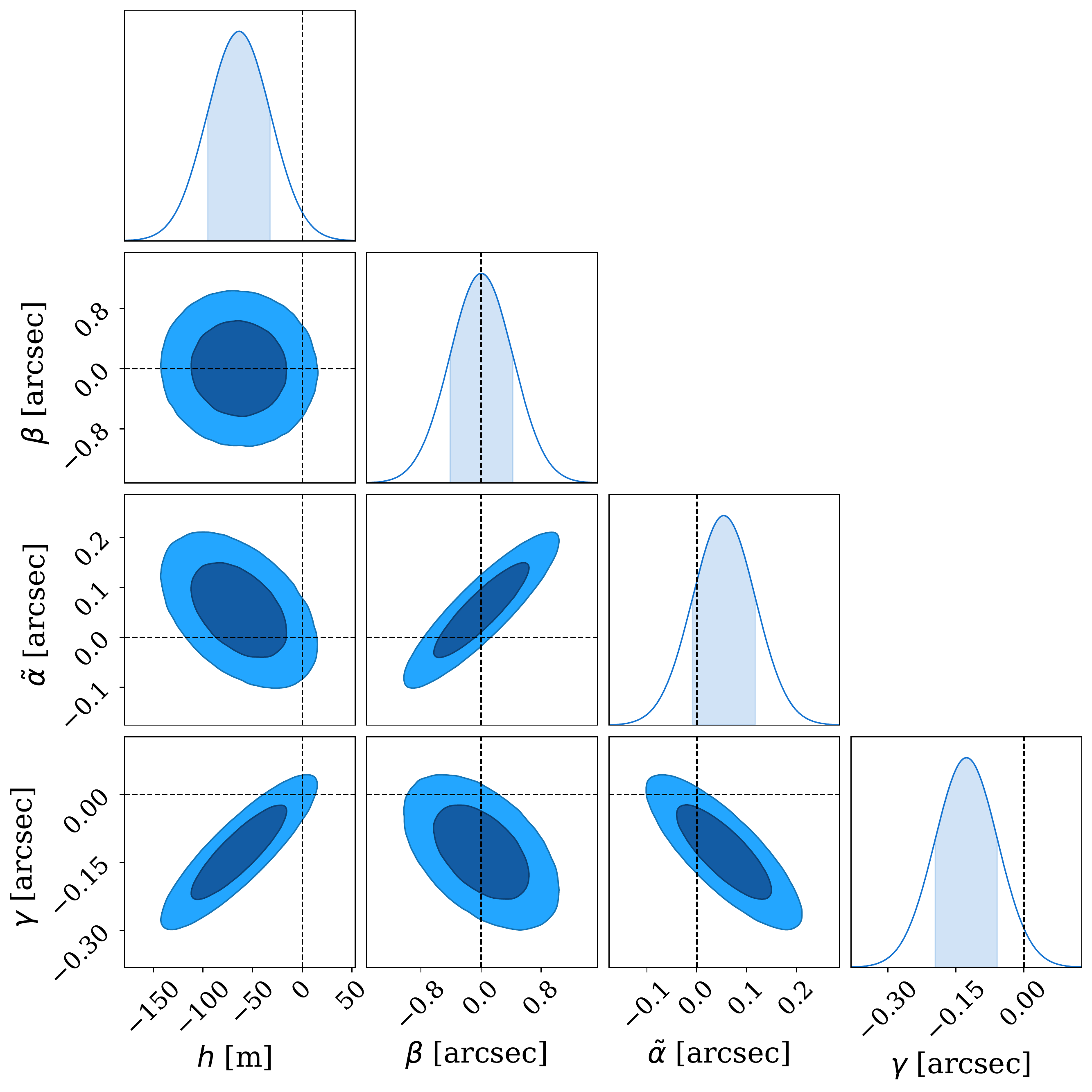}
  \caption{The combined posterior distribution of the RFI orbit parameters from 5 minutes of calibration data. Note the improvement in constraints on the angular parameters compared to the prior in Figure \ref{fig:cornerPrior} (4 orders of magnitude) and the posteriors from 10s of data in figure \ref{fig:RFIcorner3} (2 orders of magnitude). The potential estimation bias can be considered a statistical fluctuation as this disappears in when changing any aspect of the simulations. The distributions and true value have been shifted to make the true parameter values 0. This is done to make uncertainties more legible on the axes.}
  \label{fig:RFIcorner5m}
\end{figure}

Figure \ref{fig:RFIcorner3} shows a set of estimates for the RFI orbit parameters from three different data portions of the 5 minute calibration observation, as indicated in the upper right of the figure. We see that the individual estimates are of varying quality that depend on the SNR of the data used. We also see in a subset of the marginals that include the inclination parameter that the correlations across portions leading to greater constraining power when combined. Figure \ref{fig:RFIcorner5m} shows the final posterior estimate after combining the individual posterior estimates according to the equations in Section \ref{sec:comb_ind}. Table \ref{tab:RFIorbit} gives a summary of the marginal standard deviations for the priors, 10s posteriors, and 5 min. posterior estimates for the orbit parameters.

% \FloatBarrier

\begin{table}
\Large
\begin{center}
  \resizebox{0.45\textwidth}{!}{%
  \begin{tabular}{ |c|c|c|c| }
    \hline
    Parameter Name & Prior & 10s Posterior & 5 min. Posterior\\
    \hline
    \hline
    Orbit Elevation (m)                                  & 730.000   & 700.000  & 31.464        \\
    Argument of Perigee (arcsec)                         & 10106.391 & 81.932   & 0.414          \\
    Orbit Inclination (arcsec)                           & 1349.979  & 11.076   & 0.062          \\
    \begin{tabular}{@{}c@{}}Right Ascension of \\
    the Ascending Node (arcsec)\end{tabular}             & 774.384   & 6.589    & 0.068           \\
    \hline
  \end{tabular}}
  \caption{The mean marginal standard deviations for the priors and posteriors in each 10s data portion, as well as the final posterior marginal errors. The final posterior is the combination of all data portions in the 5 minute calibration observation.}
  \label{tab:RFIorbit}
\end{center}
\end{table}

% \FloatBarrier

\comment{In this section we have shown that we can not only recover accurate antenna gain estimates, but also provide rigorous uncertainty estimates from a calibration observation in the presence of a satellite-based source of RFI. Additionally, we have shown that the RFI signal itself can also be leveraged to improve our calibration constraints beyond those achieved in uncontaminated data. It should be noted however that fundamentally \tabascal\ (this method) relies on the linearity of the receivers. Unfortunately RFI signal strength can sometimes be strong enough to push the receivers into the nonlinear regime or even saturate them. The modelling of such situations is outside of the scope of this paper but can conceivably be included. We would expect however that times where the receivers are saturated would lead to irrecoverable data loss. We see \tabascal\ to be of great use when the RFI signal is not too strong and especially when it is low level where flagging algorithms do not pick it up leading to biases in standard calibration approaches.}

\subsection{Laplace Approximation vs MCMC Analysis}\label{sec:MCMC_chains}

In this section we compare the Laplace approximation with our MCMC results. We use the MCMC results as the true posterior for comparison. Later in this section, we give evidence for this claim by analyzing the MCMC chains and show their reliability as a posterior benchmark.

\begin{figure}
  \centering
  \includegraphics[width=0.45\textwidth]{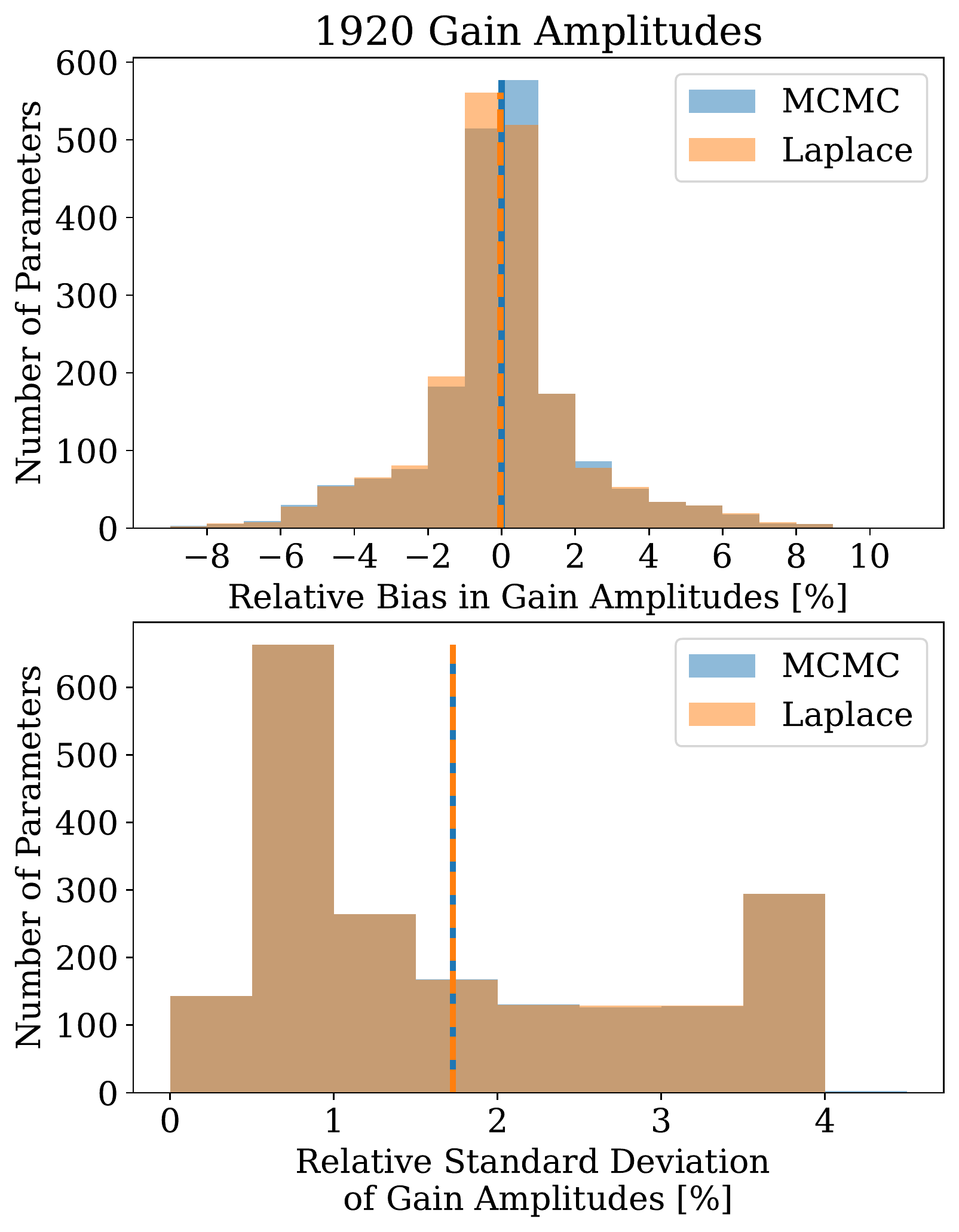}
  \caption{Comparison of biases and posterior uncertainties in our gain amplitude estimates from the Laplace approximation and MCMC. The brown region is where the distributions overlap and only in a couple bins near the centre of the top panel can we see a discrepancy. This indicates excellent agreement between the Laplace approximated posterior and the true (MCMC) posterior. The top panel shows the relative biases and the bottom panel shows the posterior standard deviations of these estimates. Each estimate is for a specific antenna and time step. The mean of each distribution is indicated by the dashed vertical line of the corresponding colour.}
  \label{fig:GampComp}
\end{figure}

\begin{figure}
  \centering
  \includegraphics[width=0.45\textwidth]{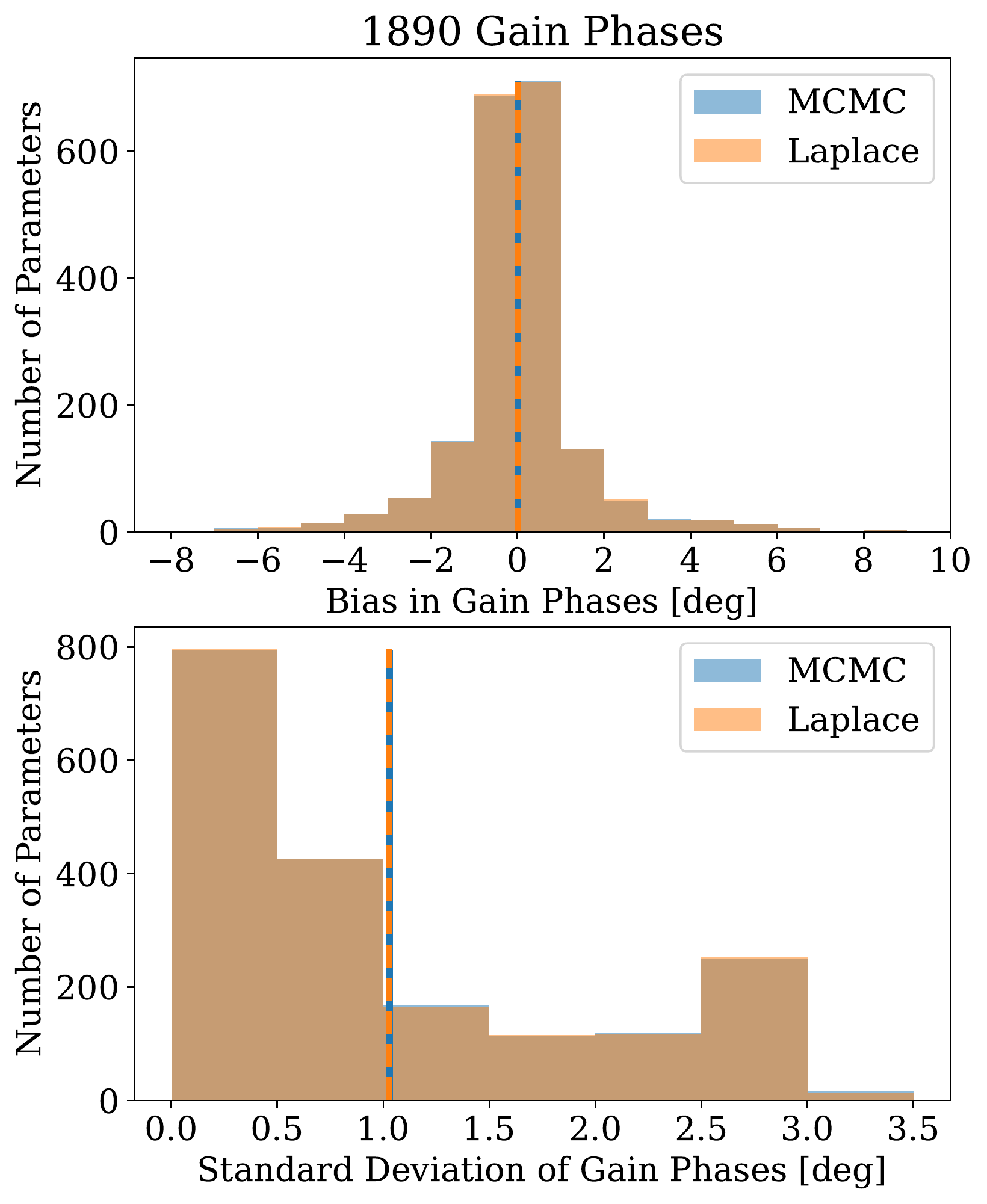}
  \caption{Comparison of biases and posterior uncertainties in our gain phase estimates from the Laplace approximation and MCMC. The distributions are indistinguishable, resulting in a brown colour as opposed to the separated blue and orange regions. This shows exceptional agreement between our Laplace approximated posterior and the true (MCMC) posterior. The top panel shows the biases in our gain phase estimates and the bottom panel shows the posterior standard deviations of these estimates. Each estimate is for a specific antenna and time step. The mean of each distribution is indicated by the dashed vertical line of the corresponding colour.}
  \label{fig:GphaseComp}
\end{figure}

Figures \ref{fig:GampComp} \& \ref{fig:GphaseComp} show the bias and posterior standard deviations on the gain parameter estimates from the Laplace approximation and MCMC. In both figures, the upper panels show the biases and the lower panels show the standard deviations. The bias tells us how accurate our best estimate is and the standard deviation is the uncertainty in the estimate. A larger bias is not necessarily a problem as long as its associated uncertainty is proportional such that the normalised bias follows the correct statistics. Figure \ref{fig:norm_bias} shows this to be true. These are the results for the 3 initial 20s portions (0-60s) of the calibration observation, described in the Section \ref{sec:method}.

As seen in the upper panel of Figure \ref{fig:GampComp}, the Laplace approximation/optimization routine shows a negligible underestimate of the gain amplitudes in comparison to MCMC. This is not present in Figure \ref{fig:norm_bias} so we can safely assume this to be a statistical fluctuation rather than a systematic effect. The posterior standard deviations, in the lower panel of Figure \ref{fig:GampComp}, overlap so well that one only sees a muddy brown color everywhere as opposed to sections with distinct blue or orange. 

% \FloatBarrier

Figure \ref{fig:GphaseComp} shows that the Laplace approximation is an excellent estimate of the posterior distribution over the gain phases. There is near perfect agreement in both biases and the posterior errors.

% \FloatBarrier

\begin{figure}
  \centering
  \includegraphics[width=0.45\textwidth]{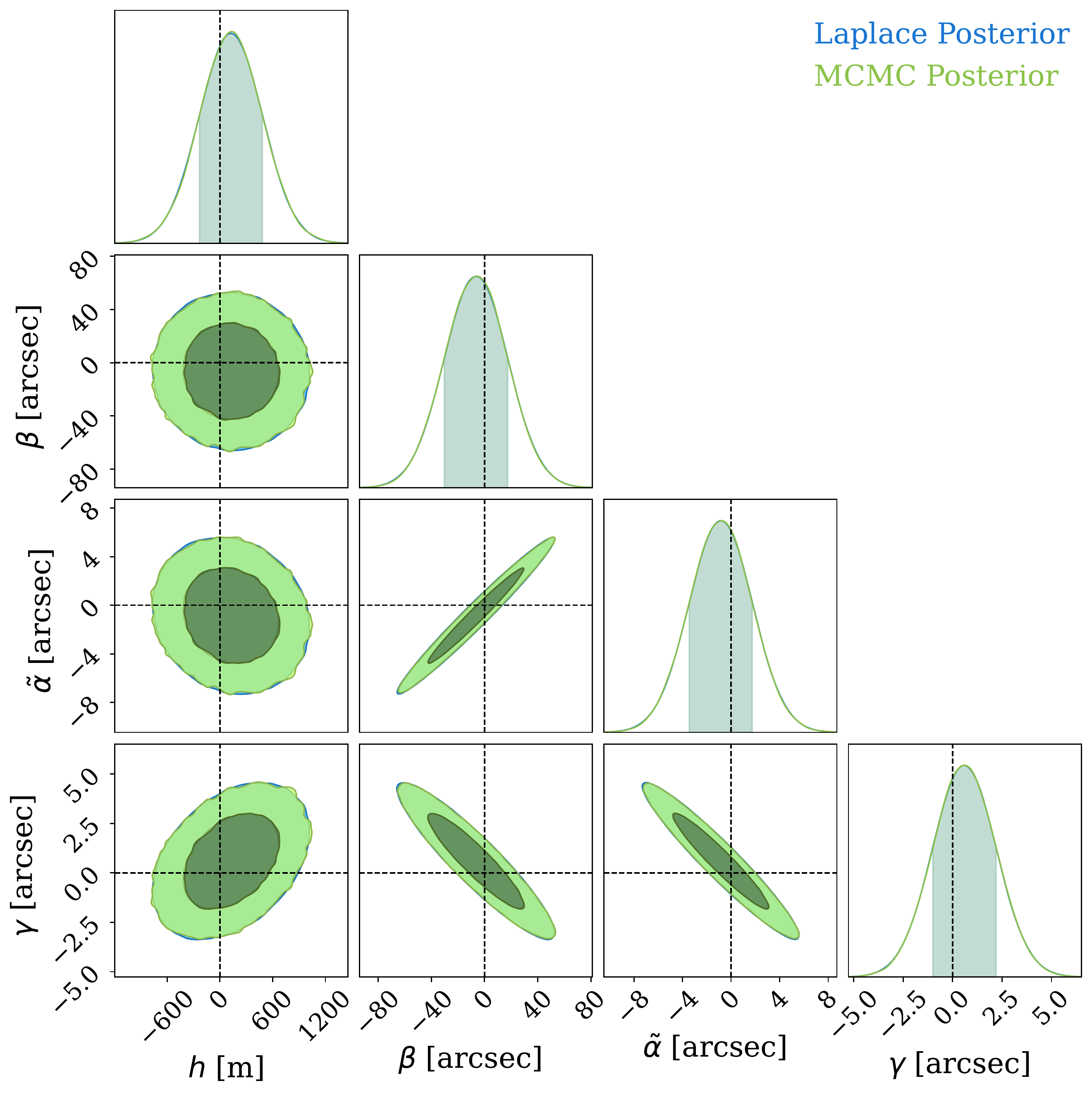}
  \caption{A corner plot of the marginalised posterior distribution of the RFI orbit parameters for a 20 second data portion. The Laplace approximated posterior is barely visible underneath the true (MCMC) posterior showing excellent agreement between the two. The 68\% and 95\% credible regions are shown for the Laplace approximated posterior (blue) and MCMC posterior (green) distributions. The distribution and true value has been shifted such that the true value is defined to be $0$.}
  \label{fig:cornerPost}
\end{figure}

Figure \ref{fig:cornerPost} shows the marginalised posterior over the RFI orbit parameters for the 0-20s portion of the calibration observation. The Laplace approximation shows excellent agreement with the true (MCMC) posterior. Only when looking very closely can one see that two distributions have been plotted.

% \FloatBarrier

Next, we analyze the convergence of our MCMC chains to gauge the reliability of the MCMC derived posterior as a benchmark. Unfortunately, no analytical solution is available for our problem, as for most real world problems, and MCMC is the gold standard for estimating the posterior distribution. An MCMC routine that follows detailed-balance and is ergodic \citep{neal2011mcmc}, as our routine does, is guaranteed to converge to the true posterior distribution in the limit of infinitely many samples. Since this is computationally intractable we must rely on `sufficiently many samples'. There are standard tools available to gauge if we have `sufficiently many samples'. The primary tool used to gauge the efficiency of an MCMC sampler is the lag-autocorrelation, denoted by $\rho_t$, of individual parameter sample chains. From this the effective sample size (ESS) of a chain, or set of chains, can be calculated. This is then used in the estimation of the MCMC standard error on individual parameters. The MCMC standard error gives an estimate of the uncertainty in the posterior standard deviation estimate, $\sigma_p$, of parameter $p$. 
\begin{equation}\label{eq:ESS}
  \sigma_{\hat{\sigma}_x} = \frac{\hat{\sigma}_x}{\sqrt{N_\text{eff}}}, \quad N_\text{eff} = \frac{N}{1+2\sum_1^\infty \rho_t}
\end{equation}
It is the uncertainty on the uncertainty. Equation \eqref{eq:ESS} gives its definition and Figure \ref{tab:MCMC_stats} shows the distributions of these, as relative errors, for our different categories of parameters. The median relative error was $\approx$0.5\% with less than 1\% of the estimates being worse than 2\%.

It is also standard to check the convergence of our chains to make sure this estimate would not get worse with more samples. The potential scale reduction factor \citep{gelman1992inference}, $\hat{R}$, also know as the Gelman-Rubin (GR) statistic, is a commonly used measure of convergence for a set of MCMC chains. The closer $\hat{R}$ is to 1 the better converged it is. In \cite{vats2021revisiting} 99\% of a random sample of papers from 2017 use an $\hat{R}$ cut-off of 1.01 or greater, the remaining 1\% used 1.003. In Table \ref{tab:MCMC_stats} we give the minimum, median, maximum and the first last percentiles of the calculated Gelman-Rubin and split Gelman-Rubin statistics. These are calculated for all estimated parameters ($\approx$6000 parameters) in the 0-60s data portions. Given that the $99^\text{th}$ Percentile values for both standard and split are below 1.01 we can confidently say that our MCMC chains are well converged. 

\def\arraystretch{1.2}%  1 is the default, change whatever you need
\begin{table}
\LARGE
\begin{center}
  \resizebox{0.45\textwidth}{!}{%
  \begin{tabular}{ |c|c|c|c|c|c| }
    \hline
           &         Minimum & $1^\text{st}$ Percentile & Median & $99^\text{th}$ Percentile & Maximum\\
    \hline
    \hline
    \begin{tabular}{@{}c@{}}Relative MCMC \\
    Standard Error \end{tabular}      & 0.183\% &  0.190\% &  0.553\% &  1.787\% &  19.144\%    \\
    \hline
    \begin{tabular}{@{}c@{}}Gelman-Rubin \\
    Statistic \end{tabular}                               & 1.0000 & 1.0000 & 1.0001 & 1.0045 & 1.0442    \\
    \hline
    \begin{tabular}{@{}c@{}}Split Gelman-Rubin \\
    Statistic \end{tabular}                         & 1.0000 & 1.00000 & 1.0001 & 1.0040 & 1.0356    \\
    \hline
  \end{tabular}}
  \caption{Distribution statistics for standard accuracy and convergence tests on MCMC posterior chains. The $99^\text{th}$ Percentile figures show excellent accuracy and convergence of our chains for $\approx$6000 parameters with only a couple of outliers.}
  \label{tab:MCMC_stats}
\end{center}
\end{table}

\section{Application to a Target Observation}\label{sec:target_obs}

\comment{In this section we present and analyze an ad hoc method to subtract RFI from the visibilities of a target observation, adjacent to the calibration observation where \tabascal\ has been used to estimate the antenna gains and the RFI orbit parameters. This stands purely to demonstrate the potential use of the estimated parameters from \tabascal\ but is not a part of the \tabascal\ method. We intend to release, in a follow up paper, a rigorous subtraction method for target observations with uncertainty estimates that fits within the \tabascal\ framework. The method presented in this section is not rigorous and does not provide uncertainty estimates yet is still surprisingly effective. Using this ad hoc method, we show the potential improvements in data retention, after flagging, and the subsequent reduction in image noise. We also demonstrate how this results in deeper source extraction with more accurate flux estimates per source. We refer to the method as RFI subtraction even though it makes use of the \tabascal\ estimated parameters.}

\begin{figure}
  \centering
  \includegraphics[width=0.45\textwidth]{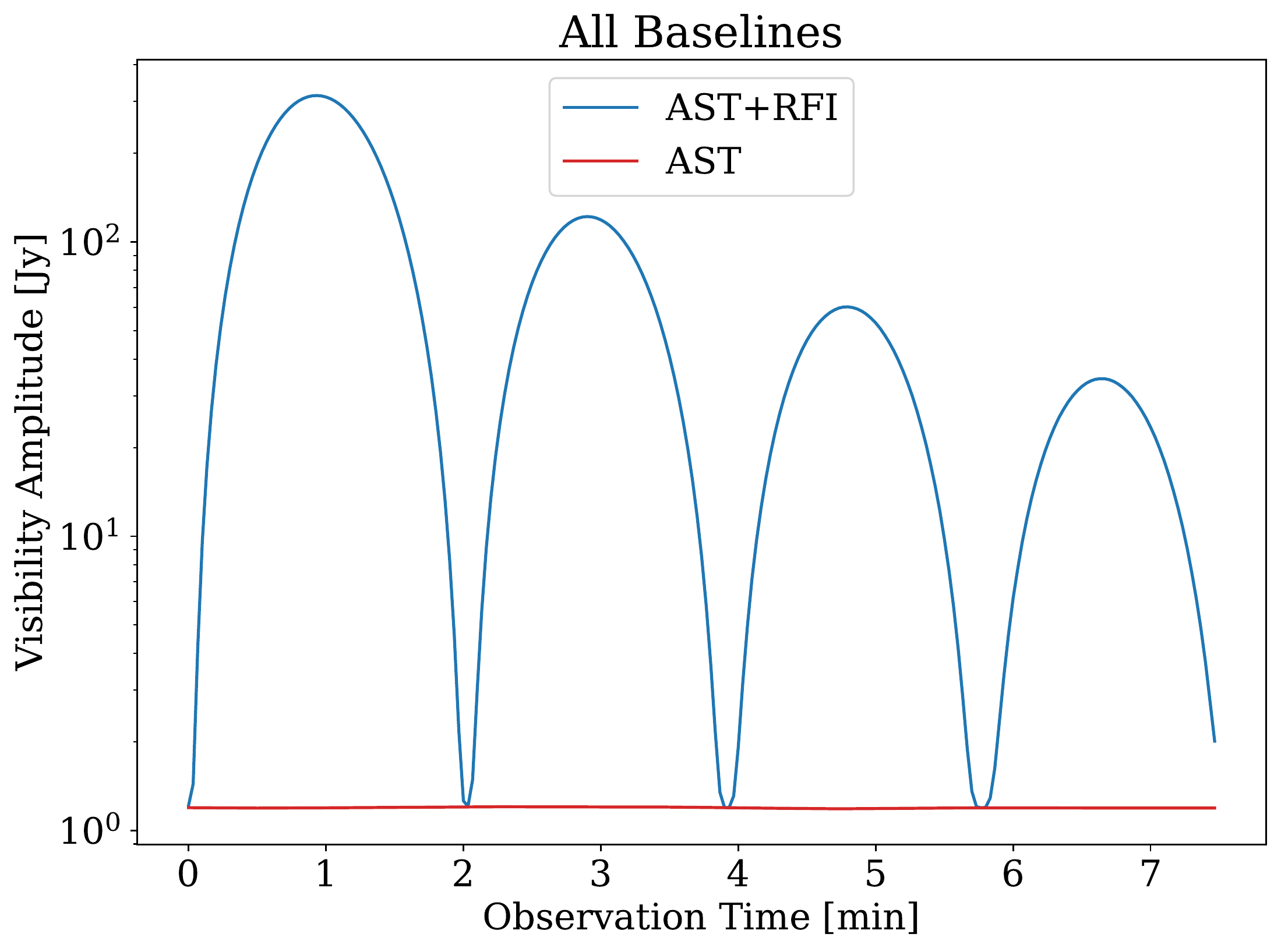}
  \caption{The visibility amplitudes in the target observation. We can see the near constant nature of the astronomical contribution (in red) in comparison to the RFI visibilities that vary by orders of magnitude over a 1 minute period. These are the average visibility magnitudes across all baselines.}
  \label{fig:target_obs_vis}
\end{figure}

To analyze this method we apply it to a 7.5 minute target observation that takes place shortly after the calibration observation. The target phase centre is $(\alpha,\delta)=(27^\circ,15^\circ)$. This position keeps the RFI location to within $10^\circ$ on the sky. The target observation has 100 uniformly distributed point sources where intensities are drawn from an exponential distribution with mean of 100 mJy. The gains used carry on from the model used for the calibration observation with the appropriate time steps and the primary beam model is identical to the calibration observation. Figure \ref{fig:target_obs_vis} shows the visibility amplitudes (averaged over baseline) for the astronomical component and the combined contaminated visibilities. No noise or gains are included to clearly show the difference in the time variability. 

% \FloatBarrier

In the basic standard approach of 1st Generation Calibration (1GC), a target observation is initially flagged for RFI followed by applying calibration solutions, from the calibration observation. After this stage various methods may be used to try and flag any remaining RFI that was missed in the first round of flagging. Hopefully, after these steps the astronomer is left with calibrated visibilities, that are free from RFI, that can go on to be imaged or as input for 2nd Generation Calibration (2GC). One of the troubles with this approach is that good calibration solutions are rarely available for RFI contaminated channels. This is due to the lack of uncontaminated data, in the calibration observation, available to calculate the gain solutions. Channels with persistent RFI may be flagged entirely for all times. Such is the case for the MeerKAT calibration pipeline on the baselines in the array core ($|(u,v)|$ < 1 km) as seen in Figure \ref{fig:rfi_lband}. This problem is solved by our method directly as we are able to accurately estimate gains in the calibration observation in the presence of  RFI. Of course having good gain solutions in contaminated channels is not enough as the the visibilities in the target observation are still contaminated. Here we describe and analyze a basic method to remove the visibility contribution from the RFI in a target observation that follows the calibration observation. This method has many similarities with those used in \cite{cotton2009low} except we predict the RFI visibility fringes from the orbit model and explicitly perform the correlator integration. We assume the RFI contribution is only from the same source that was present in the calibration observation. This demonstrates the advantage of having estimates of the parameters that describe the RFI's motion, the orbit parameters in our case.

\comment{To predict the RFI visibility component we take advantage of the expected phase wrapping caused by the movement of the RFI relative to the image frame. Since we have a good estimate of the RFI orbit parameters we can reliably predict the position of the satellite and therefore the visibility phases it would induce. The RFI visibilities are estimated in two parts; the visibility phase from the orbit parameters and then the overall amplitude over time. Using the orbit parameters, we estimate the RFI visibilities using a constant amplitude of one giving us our visibility phase. This must be done at a high enough sampling rate dictated by Equation \eqref{eq:RFI_sampling_rate}. We then calibrate the observed target visibilities using our gain estimates from the calibration observation to obtain calibrated, contaminated visibilities. Using these calibrated visibilities, we average specific baselines for which the RFI contribution fringe-washes out over time leaving, hopefully, only the astronomical (uncontaminated) visibilities. These are then subtracted from the contaminated visibilities on those specific baselines to obtain multiple noisy estimates of the RFI visibility over time. By taking the magnitude of these subtracted visibilities and averaging over the specific baselines we obtain the average RFI intensity over time that is, hopefully, representative for the entire array. To obtain the final RFI visibility estimates over time time on all baselines; we linearly interpolate the RFI intensity to sample it the same times as the RFI visibility phases, multiply them together and average back down to the data rate (2 seconds). Finally, the RFI visibilities are subtracted from the calibrated, contaminated visibilities leaving us with calibrated, astronomical (uncontaminated) visibilities.}

\begin{figure}
  \centering
  \includegraphics[width=0.45\textwidth]{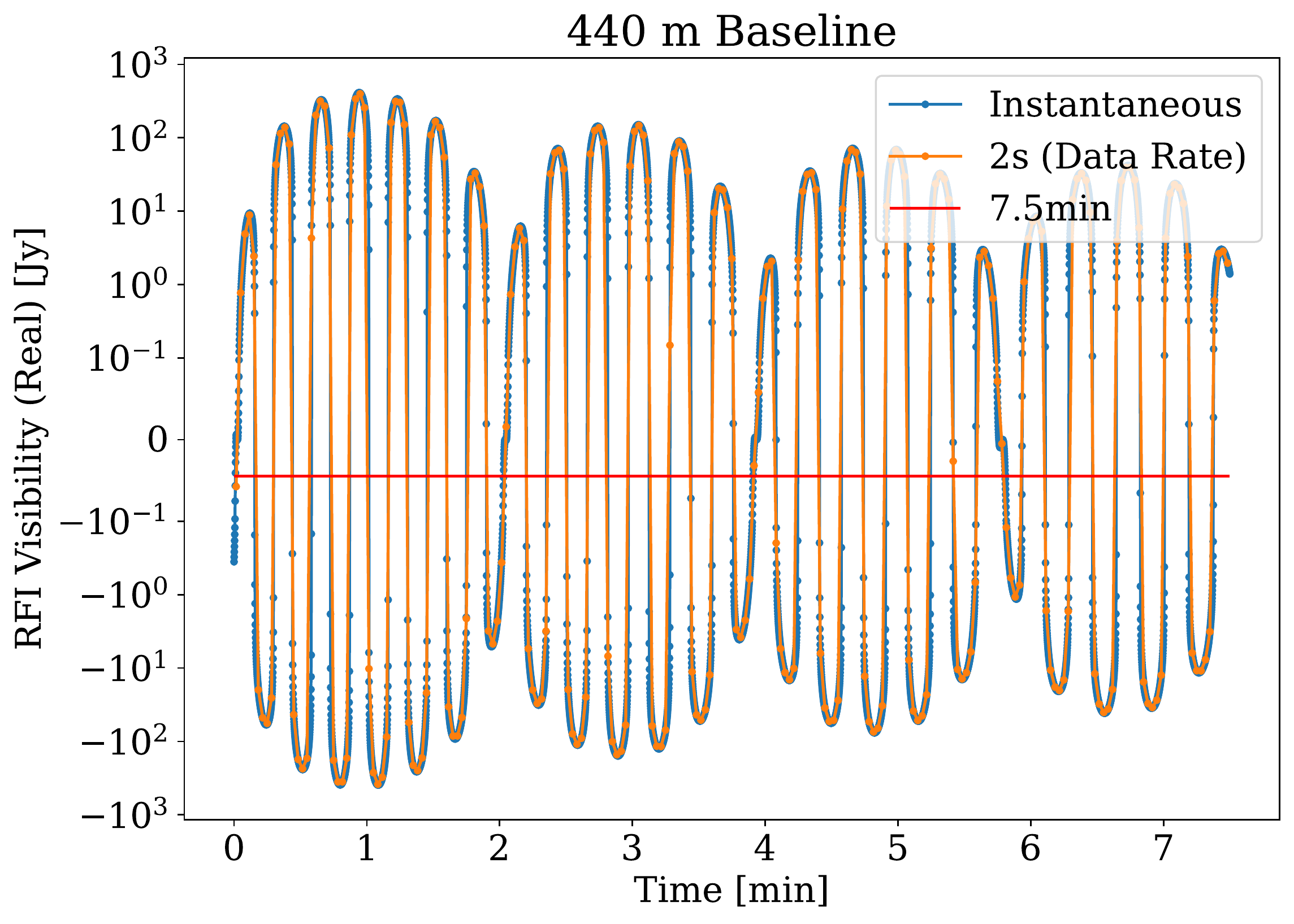}
  \caption{\comment{An example of a specifically chosen baseline used to estimate the instantaneous (blue) RFI visibility amplitude. The visibility has been averaged at different rates showing the negligible fringe-washing at the data rate (2 seconds, orange) and near complete fringe-washing at 7.5 minutes (red). In the ad hoc RFI subtraction technique presented in this section, 200 such baselines are used to estimate the RFI visibility amplitude over time.}}
  \label{fig:rfi_averaging}
\end{figure}

\comment{To obtain the best estimate for the instantaneous RFI intensity over time, the baseline over which to average should be specifically chosen. An ideal baseline is short such that the astronomical visibility deviates minimally, over the averaging time (e.g. 5 minutes) from its average but the RFI visibility fringe-washes away (averages down) maximally. Ideally, the RFI visibility has been fringe-washed minimally in each observed data point (e.g. 2 seconds). An example of one of these specific baselines, taken from the data in our analysis, where the RFI visibility has been averaged at differing rates is shown in Figure \ref{fig:rfi_averaging}. We see that the 2 second averaged (orange) RFI visibility is very close to the instantaneous visibility (blue), whereas, when averaged over 7.5 minutes (red) it is close to zero. To achieve this fine balance it is best to choose baselines on which the RFI visibility phase wraps close to an integer number of times. If the RFI intensity were constant over time (which it never is due to the primary beam modulation) we would choose baselines with as close to one phase wrap as possible. Due to the time variation of the RFI intensity, the number of phase wraps is typically best chosen to be a multiple of the number of sidelobes the source passes through. In our case from Figure \ref{fig:target_obs_vis} we can see this is four sidelobes. The specific baselines to average over are chosen to be the $N_b$ baselines on which the RFI phase wraps closest to $N_w$ integer times. The values $N_b$ and $N_w$ are chosen such that the flagging percentage, on the corrected visibilities, is minimized. We found values of $N_b=200$ and $N_w=20$ to be optimal using a grid-search.}

% \FloatBarrier

\comment{Performing this rather ad hoc procedure to remove the RFI contribution from the calibrated visibilities works surprising well but still requires some flagging. However, this is not the preferred method and was only conceived to demonstrate the potential for target observations, after estimating the gains and RFI orbit parameters from a contaminated calibration observation using \tabascal. The truly desirable solution would be to simultaneously solve for the RFI and astronomical visibilities using an optimization scheme. This will be presented in future work. Such a method would no longer require any flagging. We have deferred the presentation of such a method for a follow up paper as it deserves an in depth analysis and would make this paper too long.}

\subsection{Flagging Improvement}

In this section we perform a flagging comparison on the target observation data. For the standard case, i.e. 1GC, we have calibrated the data using the true gain solutions, averaged over time, from the calibration observation. Flagging is then performed using $\sigma$ thresholding, where $\sigma=0.65$ Jy is equal to the visibility noise, after subtracting the true astronomical visibilities. \comment{For our ad hoc RFI subtraction method we use our \tabascal\ estimated gains from the calibration observation. After this we apply the ad hoc RFI subtraction technique described in Section \ref{sec:target_obs} above.} Best or ideal refers to the use of the true gain and orbit parameters and when we reference a time it indicates the amount of calibration data that \tabascal\ used to form the estimate. \comment{For brevity we will often refer to the combination of using \tabascal\ derived parameter estimates with the previously described ad hoc RFI subtraction method for target observations simply as RFI subtraction.}

\begin{figure}
  \centering
  \includegraphics[width=0.45\textwidth]{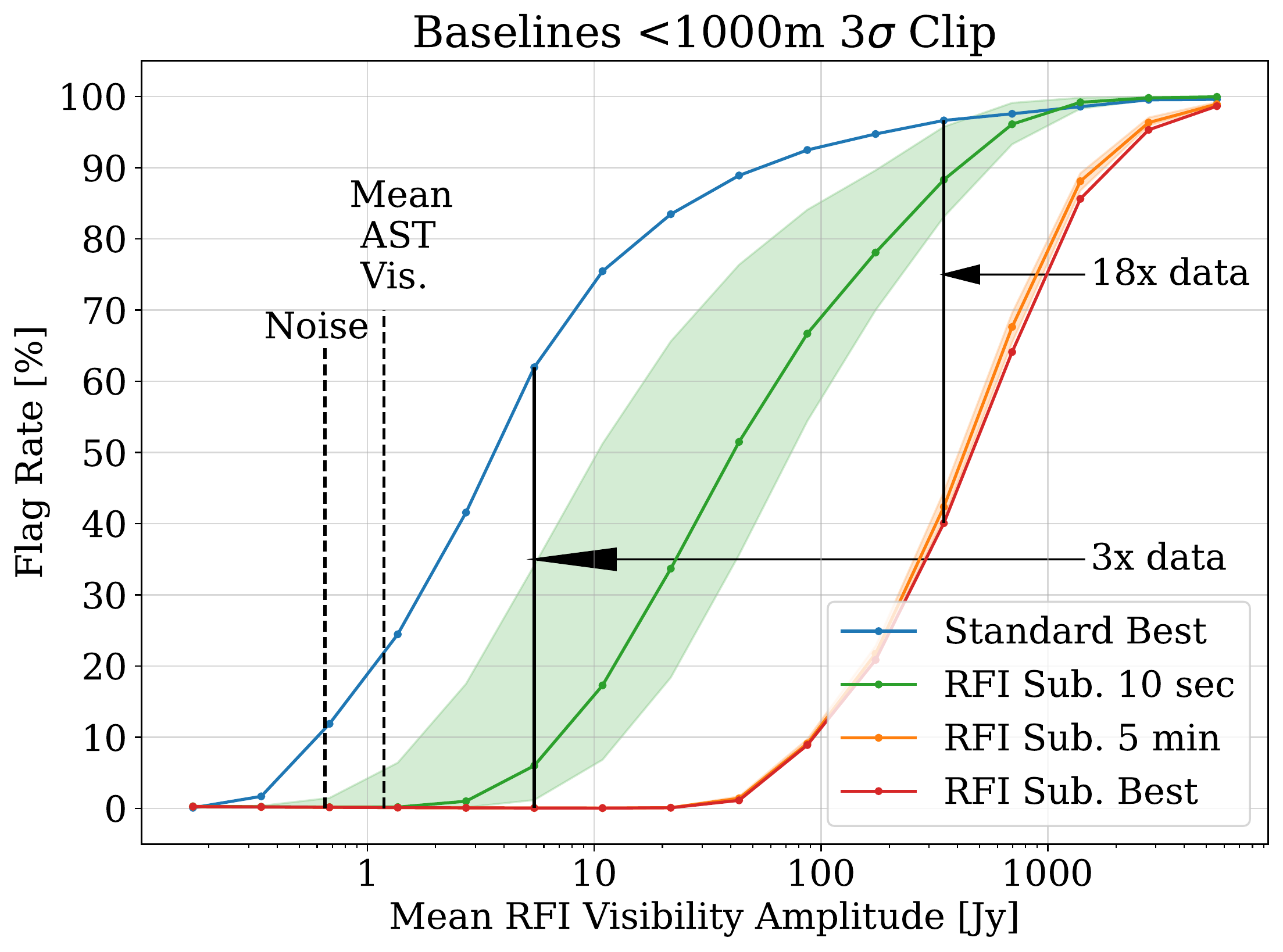}
  \caption{The percentage of data loss on baselines shorter than 1km. A dramatic improvement is seen in the 5-100 Jy region where less than 10\% data loss is observed when using RFI subtraction (red and orange) compared to a 60-90\% data loss when only performing calibration (blue). In the MeerKAT pipeline these baselines are flagged without question using a static mask. The red curve shows the performance of our method using the true gains and RFI orbit parameters. The shaded regions give the 68\% credible interval obtained from using estimates from different calibration data portions.}
  \label{fig:flagging}
\end{figure}
% \FloatBarrier

In Figure \ref{fig:flagging} we show a comparison of the percentage of data that is flagged using a $3\sigma$ flagging threshold on baselines shorter than 1km in the target observation data. We have varied the scale of the RFI visibilities in the target observation to show how our method compares when applied to data that has varying levels of RFI contamination. \comment{Due to the imperfect estimation of the RFI visibilities more flagging is required as the RFI visibility amplitude increases.} The best performance, when compared to the standard approach, is in the 5-100 Jy range leading to an approximate 70 percentage point decrease in data being flagged. This translates to approximately 3-18 times more data, in the short baselines, available for imaging. It should be noted however that this is in an optimal flagging situation where the true astronomical visibilities are known. In practice, for the case of MeerKAT, the data on these baselines would be completely flagged meaning our method is opening up the possibility of science in an untapped domain. With the effective use of the short baselines, astronomical large scale structure could now be accessed in the contaminated frequency bands. HI intensity mapping, with radio interferometers, could now probe the small scale structures in the 1-40 arcminute range at redshift of z = 0.092 - 0.235\footnote{These values are calculated using a frequency range of 1150-1300 MHz and baseline range of 30-1000 m.}.

In Figure \ref{fig:flagging} the flag rate is compared between the standard method, 1GC only, and our \comment{ad hoc RFI subtraction method.} Our method performs nearly identically in the best case scenario and using 5 minutes of \comment{\tabascal\ processed} calibration data. The confidence intervals, indicated by the shaded regions, are generated by sampling parameter sets from our \comment{\tabascal\ generated marginal posterior distributions and applying our ad hoc RFI subtraction technique} for each sample and calculating the flag rate.

The comparison in Figure \ref{fig:flagging} only takes into account the short baselines forming the core of the MeerKAT array, however, these compose over 50\% of all the baselines. Looking at all baselines the results looking remarkably similar. The improvement that \comment{RFI subtraction} brings is slightly more pronounced on shorter baselines as the RFI amplitudes are reduced on longer baselines due to the greater time smearing/phase wrapping for the RFI contribution. 

\begin{figure}
  \centering
  \includegraphics[width=0.5\textwidth]{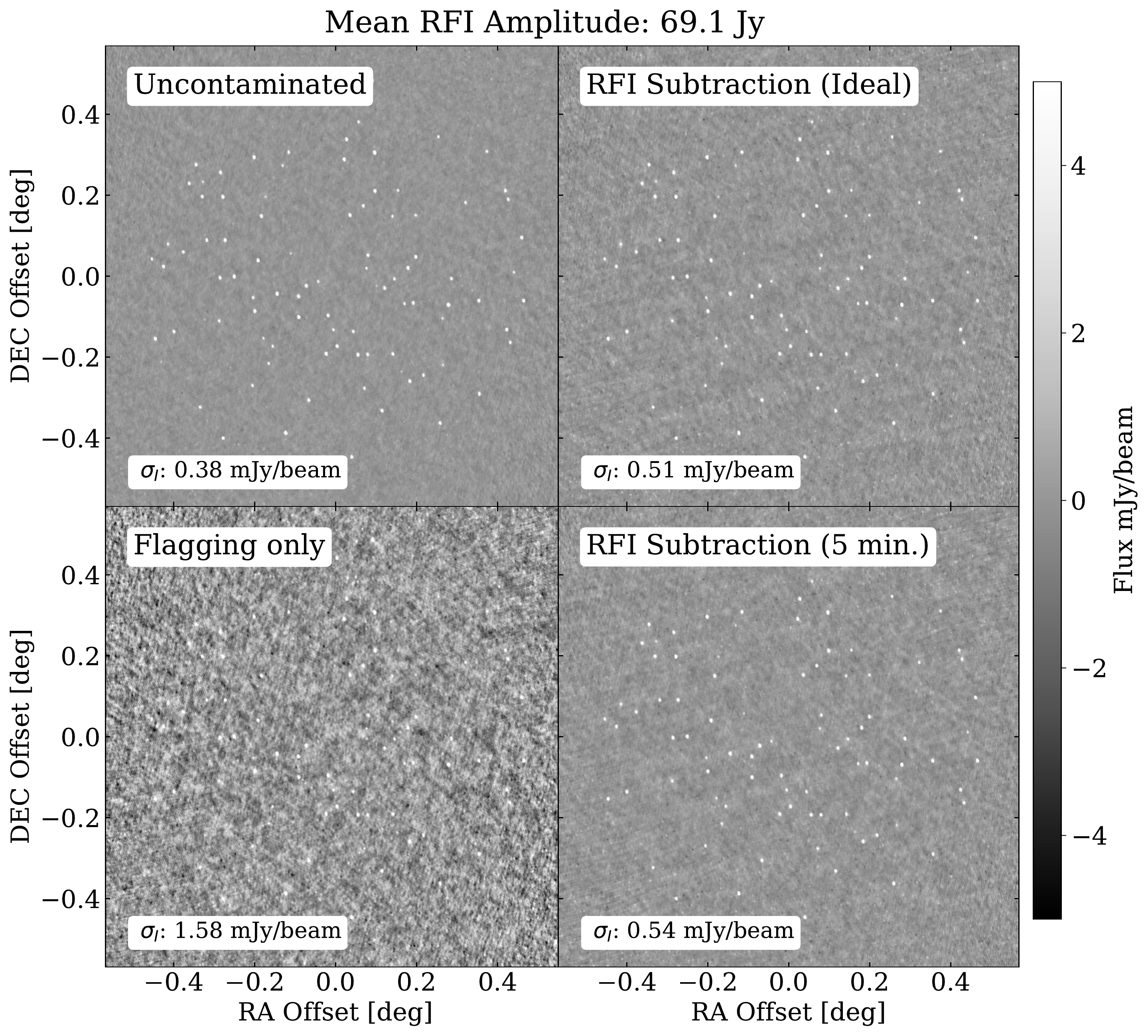}
  \caption{Imaging comparisons of a target observation of 100 point sources. RFI subtraction manages to reduce the RFI contamination significantly resulting in a third of the image noise compared to flagging alone and \comment{$2\times$} more noise than an image from uncontaminated data. Additionally, only 5 minutes of calibration data is needed to achieve near optimal results for the method. The top left image uses true astronomical visibilities with noise added. The top right image is our RFI subtraction method using the true RFI orbit and gains after 11\% of the data is flagged. The bottom left image is the standard method, 1GC only, using the true gains with 89\% of the data flagged. The bottom right image is our RFI subtraction method using the \tabascal\ estimated RFI orbit and gains from the 5 minute calibration observation. In this case 15\% of the data was flagged. For all flagging, a $3\sigma$ threshold was used.}
  \label{fig:imag_comp}
\end{figure}
% \FloatBarrier

\subsection{Imaging Comparison}

\begin{figure*}
  \centering
  \includegraphics[width=0.95\textwidth]{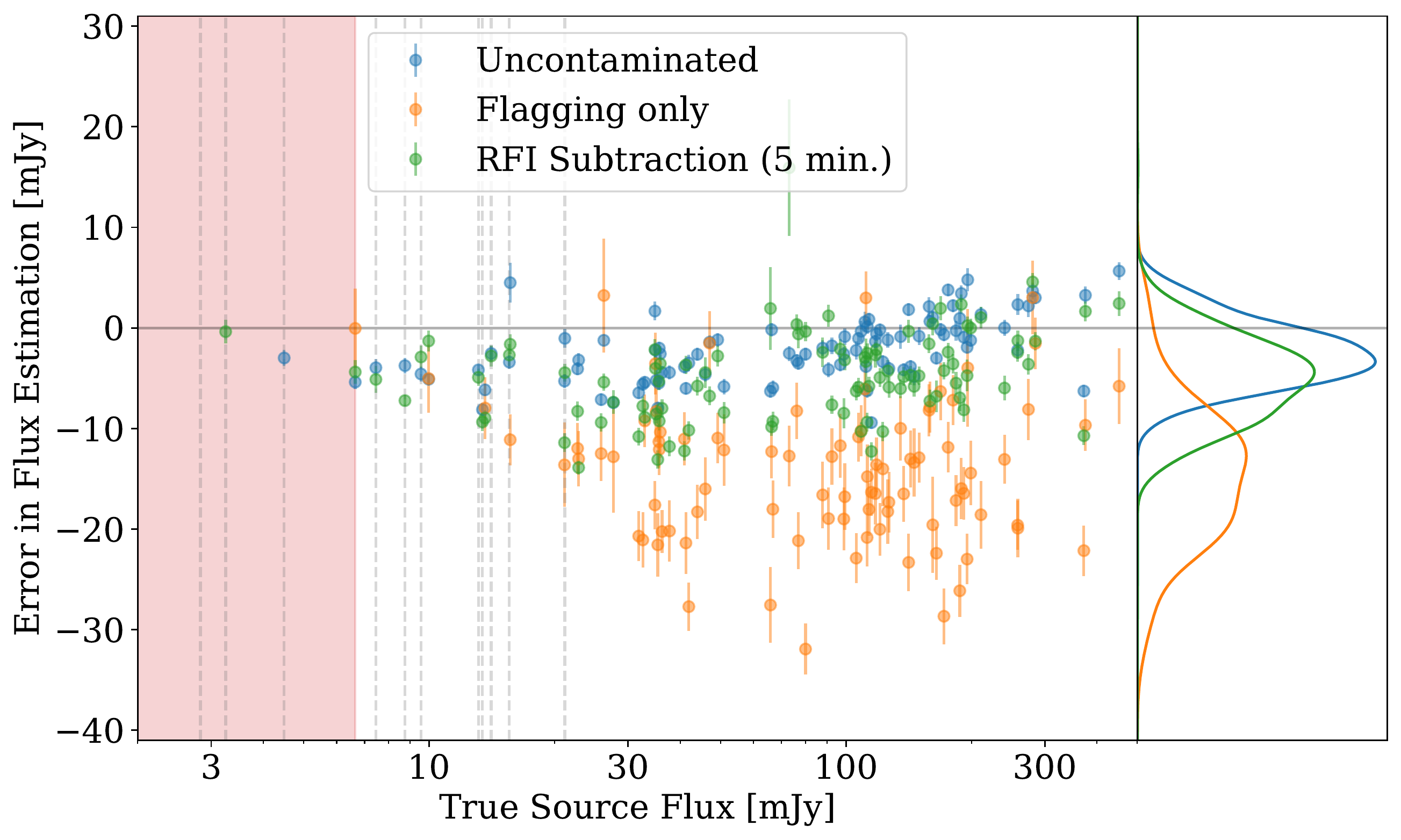}
  \caption{Flux estimation errors for sources found in the images from Figure \ref{fig:imag_comp} using \texttt{pyBDSF}. We see that making use of \tabascal\ estimates gives near optimal (uncontaminated) results while the standard approach is both biased towards lower fluxes and has larger errors. The red shaded area below about 7 mJy is where sources were completely undetected in the flagging only image. The blue markers are from the image using true astronomical visibilities with noise added. The orange markers are for the best case situation in the standard RFI flagging only approach and the green markers are for our RFI subtraction method, using \tabascal\ generated posterior means from 5 minutes of calibration data.}
  \label{fig:flux_error}
\end{figure*}
% \FloatBarrier

Imaging was performed using CASA's \texttt{tclean} \comment{in a single round with a 2 arcsecond pixel size, a Brigg's weighting scheme with robustness parameter of -0.5, and 20k iterations. We produced 2048$\times$2048 size images} with all other parameters set to default. We image four separate situations for comparison. We have an uncontaminated case, a 1GC best case using perfect calibration and only flagging, and two RFI subtraction cases where we have applied \comment{the ad hoc RFI subtraction technique and then flagging afterwards.}

The uncontaminated case uses purely astronomical visibilities with noise added. For this case we do not include any telescope response effects and no RFI contamination. We image only $V^\text{AST}_{pq} + \eta_{pq}$. For the 1GC standard case we take the observed visibilities, $V^\text{OBS}_{pq}$, and calibrate them using the true gains from the target observation. Therefore we are imaging $V^\text{OBS}_{pq}/(G_pG^*_q)$ after $3\sigma$ threshold flagging is applied. In the RFI subtraction cases, we initially perform 1GC, then we subtract an estimated RFI visibility component and finally flag the data. Flagging is done in the same way as the standard case. In our best case we use the true gains and RFI orbit parameters and in our 5 min. case we use our estimates from the 5 minute calibration observation.

Performing imaging and source extraction with lower levels of RFI amplitudes leads reduced image noise for both the standard case and RFI subtraction as would be expected and increases the number of found sources. Additionally, the bias present in source extraction for the standard (flagging only) case reduces as the RFI amplitudes decrease. As RFI amplitudes increase, the bias increases and RFI subtraction also becomes a victim of this although to a lesser degree. We find that RFI subtraction consistently performs better than flagging alone across all RFI amplitude ranges, both in bias and in image noise. 

In Figure \ref{fig:flux_error} we show a comparison on the source finding and flux recovery. For all cases we used the images in Figure \ref{fig:imag_comp}. We used \texttt{pyBDSF} with default settings to perform source finding and measurement. The same imaging and source extraction settings were used for each image. \texttt{pyBDSF} gives us source positions and fluxes all with errors among a number of other source measurements. Sources were matched to the true source model using \texttt{astropy}'s \texttt{match\_to\_catalog\_sky}. The error bars are generated by \texttt{pyBDSF}.

\comment{Only the 2 mJy source was not found by either of the ideal case or our method. Our method was comparable to the uncontaminated case finding the same number of sources. In contrast, the standard approach} only found sources down to about \comment{7} mJy. This can be attributed to the higher image noise as a result of the higher flagging rate. We see that the standard approach tends to recover less flux compared to our method and the uncontaminated case. The smoothed histograms on the right hand panel hand panel in Figure \ref{fig:flux_error} shows the distribution of flux estimation errors for each case. The distributions have been weighted by the source flux uncertainties from \texttt{pyBDSF}. We see that our method performs similarly in terms of mean error and spread compared to the uncontaminated case, whereas, the standard (flagging only) method has both a broader distribution and systematically underestimates source fluxes.

\section{Conclusions and Further Work}\label{sec:conclusions}

Calibration of radio interferometer arrays is a fundamental step in radio astronomy and is typically profoundly contaminated by Radio Frequency Interference (RFI). The usual approach to RFI is to simply cut out (flag) all obviously contaminated data, leading to significant data loss. Ideally astronomers would like to be able to effectively remove it without losing astronomical signal, but, for moving sources of RFI, this has not proven possible so far. In this paper we show that this is possible, at least for a class of RFI moving on predictable trajectories, such as satellites. The key ideas that allow this progress are (1) moving sources relative to the phase center coupled with a curved wave front (near-field) model distinguish RFI from astronomical sources, and (2) we have a good model for the trajectory of the satellite by using TLE orbit parameters.

Our algorithm, \tabascal, starts by building a forward generative model of the signal parameterized by the antenna gains, satellite orbital elements, and modulated RFI amplitudes. We then compare two approaches to estimating a posterior distribution over these parameters. The fastest method finds the best-fitting parameters (MAP) by an optimization algorithm followed by using the Laplace (Gaussian) approximation to estimate the parameter uncertainties including their covariance. The more rigorous approach uses MCMC to find the full posterior distribution without approximation. We find that the Laplace approximation works very well on our simulated data having very good agreement with the full MCMC approach. 

One of the most interesting results of our analysis is that \tabascal\ is able to calibrate using the combined astronomical + RFI signal, thus turning the contamination into an advantage to yield more precise calibration with reliable uncertainties. In application to an adjacent target observation, the \tabascal\ estimated RFI trajectory and calibration parameters can be used to estimate and subtract the RFI signal. The residual data is then flagged using sigma clipping. For a simulated MeerKAT target observation and looking at all baselines shorter than 1km we find that for a mean RFI amplitude of 17 Jy, using \tabascal\ \comment{+ an ad hoc RFI subtraction method} leads to less than $1\%$ loss of data compared to 75\% data loss from an ideal $3\sigma$ flagging algorithm. At 69 Jy the loss is $89 \%$ for the standard \comment{(flagging only)} method and 11\% for \comment{the RFI subtraction method using \tabascal\ estimates}, a nearly $9\times$ increase in data available for science. Once imaged, RFI subtracted data using \tabascal\ estimates allows recovery of faint sources that are completely missed in images from purely flagged data, i.e. the standard method. Empirically we found that using \tabascal\ \comment{+ RFI subtraction reduces the detection threshold by a factor of 3 relative to the standard method, bringing it within a factor 2 of the detection threshold for data uncontaminated by any RFI.} Furthermore, the recovered source flux distribution from \tabascal processed data was comparable with the uncontaminated data while source fluxes recovered through flagging alone were biased towards fainter fluxes.

In this work we have used \tabascal\ in only a single frequency channel. However, it is trivially applied to multiple frequencies by running it in parallel across frequency channels. \comment{We have also only included a single satellite-based RFI source however including more sources is certainly possible at the cost of more computation. We expect the computational cost to increase linearly with the number of RFI sources, time steps, frequency channels and baselines (i.e. quadratically with the number of antennas). A naive estimate for the computational cost of this approach to clean 10 GPS satellites (typical visible number at any one time) for SKA-Mid Phase 1 with $\approx 200$ antennas is $\approx 100\times$ the simulation performed in this paper. Therefore, for a 4-core laptop, as has been used here, $\approx 100$ minutes for a single frequency channel and 5 time steps (10 seconds of data sampled at 0.5 Hz). This works out to $\approx 2400$ core hours per hour of data for a single frequency channel. We expect this computational cost can be significantly improved.} 

\comment{One obvious way of achieving this speedup is the use of GPUs or other dedicated hardware for the computation, which often yield a 10-$100\times$ speedup in similar situations.  The other speedup comes when considering multiple frequency channels: clearly once we know the trajectory of a satellite we do not need to independently refit for the trajectory in other frequency channels. Similarly, in many cases the RFI amplitude will be correlated across frequency, reducing the effective number of RFI frequency parameters that need to be fit for.  The optimal manner to achieve such speedups are outside the scope of this paper and left to future work.} 

\comment{Currently \tabascal\ does not recover the astronomical signal in the target observation but instead relies on an ad hoc RFI subtraction method. This is why flagging is still required after the application of our RFI subtraction method to the target observation data.} The extension of this work is already in progress and will be presented in a follow up paper. So far we have worked with the total intensity signal, however, it is straightforward to extend this to the full polarization domain as most RFI signals are strongly polarized due to their antenna geometry.

Finally, to extend this work to real observations, it is expected that a simplified perturbation model (SGP4/SDP4) may be needed to model satellite trajectories with sufficient accuracy. These are publicly available and can be re-implemented in \JAX\ if needed.

\section*{Acknowledgements}

We thank members of the SARAO Data Science team, Radio Astronomy Research Group at SARAO and Niruj Mohan for useful discussions. CF and MK acknowledge funding by the Swiss National Science Foundation. We also acknowledge the support of the South African Radio Astronomy Observatory. This research has been conducted using resources provided by the Science and Technology Facilities Council (STFC) through the Newton Fund and SARAO. We thank Rick Perley for his insightful comments that not only improved our work but our understanding as well.

%%%%%%%%%%%%%%%%%%%%%%%%%%%%%%%%%%%%%%%%%%%%%%%%%%
\section*{Data Availability}

The data and code for recreating Figures \ref{fig:cornerPrior} to \ref{fig:flux_error} and Tables \ref{tab:RFIorbit} \& \ref{tab:MCMC_stats} are made available through the following url: \href{https://doi.org/10.5281/zenodo.7516959}{https://doi.org/10.5281/zenodo.7516959}. The code for simulating the data is available on the GitHub repository for \tabascal\ at \href{https://github.com/chrisfinlay/tabascal}{https://github.com/chrisfinlay/tabascal}. The optimization and MCMC routines, as well as, the Laplace approximation will be made available in the future on the \tabascal\ repository.

%%%%%%%%%%%%%%%%%%%% REFERENCES %%%%%%%%%%%%%%%%%%

% The best way to enter references is to use BibTeX:

\bibliographystyle{mnras}
\bibliography{main} % if your bibtex file is called example.bib

% Alternatively you could enter them by hand, like this:
% This method is tedious and prone to error if you have lots of references
%\begin{thebibliography}{99}
%\bibitem[\protect\citepauthoryear{Author}{2012}]{Author2012}
%Author A.~N., 2013, Journal of Improbable Astronomy, 1, 1
%\bibitem[\protect\citepauthoryear{Others}{2013}]{Others2013}
%Others S., 2012, Journal of Interesting Stuff, 17, 198
%\end{thebibliography}

%%%%%%%%%%%%%%%%%%%%%%%%%%%%%%%%%%%%%%%%%%%%%%%%%%

%%%%%%%%%%%%%%%%% APPENDICES %%%%%%%%%%%%%%%%%%%%%

\appendix

\section{Combining Measurements}

\subsection{Correlation Structure in the Prior}\label{sec:rfi_corr}

By using a correlated prior for the modulated RFI amplitudes one is able to improve the constraints on the gain amplitude estimates. this is because there is a pathway for information to be shared between antennas much like when we assume the astronomical source brightness is the same for all antennas. For our prior on the modulated RFI amplitudes across antennas, at each time step, we can assume the amplitudes are drawn from a multivariate normal distribution with mean 0. We then set a prior covariance, that includes correlation across antennas, for each time step. We do this as we expect the RFI amplitudes, across antennas, within a single time step to be very similar to one another with the assumptions that the primary beam patterns of the antennas are similar. We formulate the covariance matrix using a covariance function with a squared exponential kernel in the same way as a Gaussian process. We use the same covariance at each time step. Equation \eqref{eq:rfi_amp} gives the prior distribution of the modulated RFI amplitudes, where $\bm{A}^\text{RFI}(t_j)$ is the vector of amplitudes at time step $t_j$ and $\bm{0}$ is the zero vector. The covariance, $\Sigma_A$, of the prior distribution is defined in Equation \eqref{eq:rfi_cov} where $\vec{x}_p$ is the position of antennas $p$. For example, with the variance, $\sigma^2_A$, and the length scale, $l_A$, of the covariance function set to 10 000 Jy and 10 km respectively, we allow a large variation in amplitude but impose strong correlations across the entire antenna array.

\begin{equation}\label{eq:rfi_amp}
  \bm{A}^\text{RFI}(t_j) \sim \mathcal{N}(\bm{0},\Sigma_A)
\end{equation}

\begin{equation}\label{eq:rfi_cov}
  \left(\Sigma_A \right)_{pq} \left( \sigma_A^2,l_A \right) = \sigma_A^2 \exp \left[ - \frac{|\vec{x}_p-\vec{x}_q|^2}{2l_A^2} \right]
\end{equation}

\subsection{Independent Measurements}\label{sec:comb_ind}

Let us have $N$ independent measurements of an $m$-dimensional variable $X$. Each measurement, $x_n$, is normally distributed with mean $\mu_{x_n}$ and covariance $C_{x_n}$. The combination of all N measurements is simply the normalised product of their probability densities. Since each measurement is normally distributed the combined probability density will also be a normal distribution. Note that $n$ is used to index individual measurements.

\begin{equation}
  x_n \sim \mathcal{N}(\mu_{x_n}, C_{x_n})
\end{equation}

\begin{equation}
  x \sim \mathcal{N}(\mu_x, C_x) = \prod_n \mathcal{N}\left( \mu_{x_n}, C_{x_n} \right)
\end{equation}

\begin{equation}\label{eq:combine_cov}
  C_x = \left( \sum_n C_{x_n}^{-1} \right) ^{-1}
\end{equation}

\begin{equation}\label{eq:combine_mean}
  \mu_x = C_x \left( \sum_n C_{x_n}^{-1}\mu_{x_n} \right)
\end{equation}

\subsection{Correlated Measurements}\label{sec:comb_corr}

Let us have $N$ correlated measurements of a 1-dimensional variable. Each measurement is denoted by $x_j$ and the total error covariance of all measurements is $C_x$, of dimension $N \times N$. The combined measurement $\bar{x}$ is calculated using Equation \eqref{eq:xbar} with its associated error variance $\sigma^2_{\bar{x}}$ as calculated using Equation \eqref{eq:varxbar}. Note the $j,k$ and $l$ subscripts are used for indices of associated measurement vector and error matrices. The following equations are taken from \cite{avery1996combining}.

$\bar{x}$

\begin{equation}\label{eq:xbar}
  \bar{x} = \sum_j w_j x_j
\end{equation}

\begin{equation}\label{eq:varxbar}
  \sigma^2_{\bar{x}} = \sum_{jk} w_j w_k \left( C_x \right)_{jk}
\end{equation}

where $w_j$ is defined by Equation \eqref{eq:weights} below

\begin{equation}\label{eq:weights}
  w_j = \frac{\sum_k \left( C_x^{-1} \right)_{jk}}{\sum_{kl} \left( C_x^{-1} \right)_{kl}}
\end{equation}

\subsection{Independent 1-D Measurements}

To check our solutions for both Sections \ref{sec:comb_ind} \& \ref{sec:comb_corr} we can consider a set of independent 1-dimensional measurements. In this case our measurement are denoted by $x_j$ and our error covariances become $(C_x)_{jk} = \delta_{jk}\sigma^2_j$ and $(C_{x_j}) = \sigma^2_j$ where $\sigma^2_j$ are the individual error variances for each measurement $x_j$. The combined measurement $\bar{x}$ with error variance
$\sigma^2_{\bar{x}}$ is calculated using Equations \eqref{eq:1d_ind}

\begin{align}\label{eq:1d_ind}
  \bar{x} &= \sigma^2_{\bar{x}} \sum_j \sigma^{-2}_jx_j \\
  \sigma^2_{\bar{x}} &= \frac{1}{\sum_j \sigma^{-2}_j}
\end{align}

%%%%%%%%%%%%%%%%%%%%%%%%%%%%%%%%%%%%%%%%%%%%%%%%%%

% Don't change these lines
\bsp	% typesetting comment
\label{lastpage}
\end{document}